\RequirePackage{fix-cm}
\documentclass[smallcondensed]{svjour3}     
\smartqed  
\usepackage{graphicx}
\usepackage[misc,geometry]{ifsym} 
\usepackage{lineno,hyperref}
\usepackage{color}
\usepackage{amsfonts}
\usepackage{amsmath}
\usepackage{amssymb}
\usepackage{subfigure}

\makeatletter
\def\@xfootnote[#1]{%
  \protected@xdef\@thefnmark{#1}%
  \@footnotemark\@footnotetext}
\makeatother

\newcommand{\bff}{\mathbf{f}}

\newcommand{\bfx}{\mathbf{x}}

\newcommand{\bfz}{\mathbf{z}}
\newcommand{\bfP}{\mathbf{P}}
\newcommand{\bfC}{\mathbf{C}}

\newcommand{\bfG}{\mathbf{G}}

\newcommand{\bfT}{\mathbf{T}}

\newcommand{\bfZ}{\mathbf{Z}}

\newcommand{\mm}[1]{\rm mm}

\newcommand{\vol}{{\cal V}}

\newcommand{\mbf}[1]{\mathbf{#1}}
\newcommand{\mcal}[1]{\mathcal{#1}}

\newcommand{\beq}{\begin{equation}}
\newcommand{\eeq}{\end{equation}}
\newcommand{\bea}{\begin{eqnarray}}
\newcommand{\eea}{\end{eqnarray}}

\newcommand{\dx}{\Delta x}

\newcommand{\iph}{{i+\frac{1}{2}}}
\newcommand{\imh}{{i-\frac{1}{2}}}

\newcommand{\bit}{\begin{itemize}}
\newcommand{\eit}{\end{itemize}}
\newcommand{\ben}{\begin{enumerate}}
\newcommand{\een}{\end{enumerate}}

\newcommand{\bA}{\mathbf{A}}

\newcommand{\bC}{\mathbf{C}}

\newcommand{\bF}{\mathbf{F}}
\newcommand{\bG}{\mathbf{G}}

\newcommand{\bI}{\mathbf{I}}
\newcommand{\bK}{\mathbf{K}}

\newcommand{\bM}{\mathbf{M}}

\newcommand{\bT}{\mathbf{T}}

\newcommand{\be}{\mathbf{e}}

\newcommand{\buu}{\mathbf{u}}
\newcommand{\bvv}{\mathbf{v}}
\newcommand{\bx}{\mathbf{x}}
\newcommand{\by}{\mathbf{y}}
\newcommand{\bz}{\mathbf{z}}

\newcommand{\inv}{^{-1}}
\newcommand{\iter}[1]{^{(#1)}}

\newcommand{\erf}[1]{\operatorname{erf} \left [ #1 \right ]}
\newcommand{\expo}[1]{\exp \left [ #1 \right ]}

\begin{document}

\title{A New Class of High-Order Methods for Fluid Dynamics Simulations using Gaussian Process Modeling
\thanks{This work was supported in part at the University of Chicago by 
the U.S. Department of Energy (DOE) under contract B523820 
to the NNSA ASC/Alliances Center for Astrophysical Thermonuclear Flashes; 
the U.S. DOE NNSA ASC through the Argonne Institute for Computing in Science 
under field work proposal 57789; 
and the National Science Foundation under grant AST-0909132.}
}

\titlerunning{High-order Gaussian Process Methods for CFD}        

\author{Adam Reyes      \and
            Dongwook Lee  \and
            Carlo Graziani \and
            Petros Tzeferacos
}


\institute{Adam Reyes \at
              Department of Physics, University of California, Santa Cruz, CA, U.S.A \\
              \email{acreyes@ucsc.edu}
	  %
           \and
           Dongwook Lee (\Letter)  \at
           Applied Mathematics and Statistics, University of California, Santa Cruz, CA, U.S.A \\
           \email{dlee79@ucsc.edu} 
           \and
           Carlo Graziani \at
           Flash Center for Computational Science, Department of Astronomy \& Astrophysics, University of Chicago, IL, U.S.A \\
           \email{carlo@oddjob.uchicago.edu}
           \and
           Petros Tzeferacos \at
           Flash Center for Computational Science, Department of Astronomy \& Astrophysics, University of Chicago, IL, U.S.A; 
           Physics, Oxford University, U.K\\
           \email{petros.tzeferacos@flash.uchicago.edu}  
}

\date{Received: date / Accepted: date}

\maketitle

\begin{abstract}
We introduce an entirely new class of high-order methods for computational fluid dynamics (CFD) based on
the Gaussian Process (GP) family of stochastic functions.
Our approach is to use kernel-based GP prediction methods to
interpolate/reconstruct high-order approximations for solving hyperbolic PDEs. 
We present the GP approach as a new formulation of high-order
(magneto)hydrodynamic state variable interpolation 
that furnishes an alternative to conventional polynomial-based approaches.
\keywords{Gaussian Processes \and
stochastic models \and 
high-order methods \and 
finite volume method \and 
gas dynamics \and
magnetohydrodynamics}
\end{abstract}

\bibliographystyle{spmpsci}      

\section{Introduction}
\label{sec:intro}
Cutting edge simulations of gas dynamics and magnetohydrodynamics (MHD) 
have been among the headliner applications of scientific 
high-performance computing (HPC) 
\cite{Dongarra2012future,Dongarra2010recent,Keyes2013multiphysics,Subcommittee2014top}.
They are expected to remain important as 
new HPC architectures of ever more powerful capabilities come online in the 
decades to come.  A notable trend in recent HPC developments concerns
the hardware design of the newer architectures: 
It is expected that newer HPC architectures will feature a radical 
change in the balance between computation and memory resources, with 
memory per compute core declining dramatically from current levels
\cite{Attig2011,Dongarra2012future,Subcommittee2014top}.
This trend tells us that new
algorithmic strategies will be required to meet the goals of saving memory 
and accommodating increased computation.
This paradigm shift in designing scientific algorithms 
has become a great import in HPC applications.  
In the context of numerical methods for computational fluid dynamics (CFD), 
one desirable approach is to design high-order accurate methods \cite{Subcommittee2014top} that, 
in contrast to low-order methods, can
achieve an increased target solution accuracy more efficiently and quickly
by computing increased higher-order floating-point approximations
on a given grid resolution \cite{Hesthaven2007,LeVeque2002,leveque2007finite}. 
This approach embodies in a concrete manner the desired tradeoff 
between memory and computation by exercising more computation per memory
 -- or equivalently, the equal amount of computation with less memory.

Within the broad framework of finite difference method (FDM) and finite volume method (FVM) discretizations,
discrete algorithms of data interpolation and reconstruction play 
a key role in numerical methods for PDE integration \cite{LeVeque2002,leveque2007finite,toro2009}.
They are frequently the limiting factor in the convergence rate, 
efficiency, and algorithmic complexity of a numerical scheme.  
The general procedure in 1D high-order conservative FDM is 
to pursue high-order approximations of flux function values ${\hat{F}_\iph}$
at interfaces, by interpolating the set of
the interface flux function values $\{F(q_{i-p}), \dots, F(q_{i+r}) \}$, 
each of which is evaluated as pointwise value at $q_{k}, k=i-p, \dots, i+r$, for some integers $p$ and $r$.
Mathematically, this is formulated as 
\beq
\hat{F}_\iph=\mcal{I}\Big(F(q_{i-p}), \dots, F(q_{i+r})\Big),
\eeq 
where $\mcal{I}(\cdot)$ is a highly accurate
interpolation scheme providing stable numerical interface flux values that are evaluated as pointwise values of $q_k$ over a
stencil of interpolation, 
$\cup_{k=i-p}^{i+r}I_k$, where $I_k$ denotes the $k$-th cell (e.g., $[x_{i-1/2}, x_{i+1/2}]$ in 1D)
\cite{jiang1996efficient,Mignone2010b}.
By contrast, the procedure of 1D high-order FVM begins with a set of the cell volume-averaged values
\beq
\langle q_i\rangle=\frac{1}{\dx}\int_{\dx}q(x,t^n)dx
\eeq 
as initial conditions, 
and seeks a pair of high-order accurate reconstructed pointwise
Riemann state values 
\beq
({q}_\iph^L, {q}_\iph^R)=\mcal{R}\Big(\langle q_{i-p}\rangle, \dots, \langle q_{i+r}\rangle\Big)
\eeq
at the cell interfaces $x_\iph$
using a high-order reconstruction scheme $\mcal{R}(\cdot)$ over the stencil of reconstruction, $\cup_{k=i-p}^{i+r}I_k$. 
High-order FV fluxes ${\hat{F}_\iph}$ are then evaluated by solving Riemann problems at the interfaces $x_\iph$
using the Riemann state pair $({q}_\iph^L, {q}_\iph^R)$ as inputs
\cite{buchmuller2014improved,lee2013solution,LeVeque2002,mccorquodale2011high,shu2009high,toro2009}.

More generally, interpolation and reconstruction 
are not only essential 
for estimating high-order accurate approximations for
fluxes 
at quadrature points on each cell, 
but also for interface tracking; 
for prolonging states from 
coarse zones to corresponding refined zones in adaptive-mesh refinement 
(AMR) schemes; and for various other contexts associated with high-order solutions.  
In CFD simulations, these interpolation and reconstruction algorithms must be 
carried out as accurately as possible, because, to a large extent, their accuracy is one of the key factors 
that determines the overall accuracy of the simulation.

Polynomial-based approaches are the most successful and popular 
among interpolation/reconstruction 
methods in this field. 
There are a couple of convincing reasons for this state of affairs.  
First, they are easily relatable to Taylor expansion, the  
most familiar of function approximations.  Second, the nominal
$N$-th order accuracy of polynomial interpolation/reconstruction is derived from
using polynomials of degree $(N-1)$, 
bearing a leading term of the error that scales with
$\mathcal{O}(\Delta^{N})$ as the local grid spacing $\Delta$ approaches to zero
\cite{LeVeque2002,leveque2007finite,toro2009}.
However, the simplicity of polynomial interpolation/reconstruction comes at a price:  
The polynomial approach is notoriously prone to oscillations in data fitting, 
especially with discontinuous data  \cite{Gottlieb1997}; Furthermore,
in many practical situations
the high-order polynomial interpolation/reconstruction
must be carried out on a {\it fixed} size of stencils,
whereby there is a one-to-one relationship between the order of the interpolation/reconstruction
and the size of the stencils. This becomes a restriction in particular when unstructured meshes
are considered in multiple spatial dimensions.
Lastly, another related major issue lies in the fact that the algorithmic complexity of 
such polynomial based schemes typically grows with order of accuracy \cite{Gerolymos2009},
as well as with spatial dimensionality \cite{buchmuller2014improved,mccorquodale2011high,zhang2011order} in FVM.

To overcome the aforementioned issues in polynomial methods, practitioners have developed
in the last decades
the so-called ``non-polynomial'' interpolation/reconstruction based on
the mesh-free Radial Basis Function (RBF) approximations.
The core idea is to replace
the polynomial interpolants with RBFs, 
which is a part of a very general class of 
approximants from the field known as \emph{Optimal 
Recovery} (OR) \cite{wendland2010}.
Several interpolation techniques in OR have shown practicable
in the framework of solving
hyperbolic PDEs \cite{Katz2009,morton2007,sonar1996},
parabolic PDEs \cite{moroney2006,moroney2007}, 
diffusion and reaction-diffusion PDEs \cite{shankar2015radial},
boundary value problems of elliptic PDEs \cite{liu2015kansa},
interpolations on irregular domains \cite{chen2016reduced,heryudono2010radial,martel2016stability}, 
and also interpolations on a more general set of scattered data \cite{Franke1982}.
Historically, RBFs were introduced to seek exact function interpolations \cite{powell1985radial}.
Recently, the RBF approximations have been combined with the 
key ideas of handling discontinuities in the 
ENO \cite{harten1987uniformly} and WENO \cite{jiang1996efficient,liu1994weighted} methods.
Such approaches, termed as ENO/WENO-RBF, have been extended to 
solve nonlinear scalar equations and the Euler equations 
in the FDM \cite{jung2010recovery} and the FVM \cite{guo2017rbf} frameworks.
These studies focused on
designing their RBF methods with the use of adaptive shape parameters to control local errors.
Also in \cite{bigoni2016adaptive}, two types of multiquadrics and polyharmonic spline RBFs 
were used to model the Euler equations with a strategy of selecting optimal shape parameters for 
different RBF orders.
Stability analysis on the fully discretized hyperbolic PDEs in both space and time using the multiquadrics RBF
is reported in \cite{chen2012matrix}.
While there exist a few conceptual resemblances between these RBF approaches and our new GP method, 
the fundamental differences that distinguish the two approaches are discussed later in this paper.

The goal in this article is to develop a new class of high-order methods that
overcomes the aforementioned difficulties in polynomial approaches
by exploiting the alternative perspective afforded by GP modeling,
a methodology borrowed from the field of statistical data modeling. 
In view of the novelty of our approach, which
bridges the two distinct research fields of statistical modeling and CFD,
it is our intention in this paper to first construct a mathematical
formulation in 1D framework. The current study will serve as a theoretical foundation
for later multidimensional extensions of the GP modeling for CFD applications, topics of which
will be studied in our future work.

The GP method is a class of high-order 
schemes primarily designed  
for numerical evolution of hyperbolic PDEs, $q_t + \nabla \cdot F(q) = 0$.
To this end, we describe the new high-order Gaussian Process (GP) approximation strategies in two steps:
\ben
\item GP {\it{interpolation}} that works on pointwise values of $q(x_i)$ as both inputs and outputs, and
\item GP {\it{reconstruction}} that works on volume-averaged values $\langle q_i\rangle=\frac{1}{\dx}\int_{\dx}q(x,t^n)dx$
as inputs, reconstructing pointwise values as outputs.
\een
GP interpolation will provide a ``baseline'' formulation of using GP as a new high-order interpolator
operating on the \textit{same} data type,
while GP reconstruction will serve as a high-order reconstructor operating on two \textit{different} types of data.

\section{Gaussian Process Modeling}
\label{sec:Gaussian-processes}
The theory of GP, and more generally of stochastic 
functions, dates back to the work of Wiener \cite{wiener1949} and Kolmogorov \cite{kolmogorov1941}.  Modern-day applications 
are numerous: Just in the physical sciences, GP prediction is in common use 
in meteorology, geology, and time-series analysis 
\cite{bishop2007pattern,rasmussen2005,wahba1995}, and in cosmology, where GP models furnish 
the standard description of the Cosmic Microwave Background \cite{bond1999}. 
Applications abound in many other fields, in particular wherever spatial or 
time-series data requires 
``nonparametric'' modeling \cite{bishop2007pattern,rasmussen2005,stein1999}.
Within the perspective of CFD applications,
our goal in this study is to use predictive GP modeling 
that is processed by training observed data, (e.g., cell-averaged fluid variables at cell centers) 
to produce a ``data-informed" prediction (e.g., pointwise Riemann state values at cell interfaces).
In what follows we give a brief overview on GP from statistical perspective (Section \ref{sec:gp_stat}), 
followed by our strategies of tuning 
GP for high-order interpolation (Section \ref{sec:gp_interpolation}) 
and reconstruction (Section \ref{sec:gp_reconstruction}) in CFD applications.
Readers who wish to pursue the subject in greater detail are referred to 
\cite{bishop2007pattern,rasmussen2005,stein1999}.

\subsection{GP -- Statistical Perspective}
\label{sec:gp_stat}
GP is a class of stochastic processes, i.e., processes that sample functions 
(rather than points) from an infinite dimensional function space. Initially 
one specifies a so-called \textit{prior probability distribution} over the 
function space. Then, given a sample of function values at some set of points, 
one ``trains'' the model by regarding the sample as data and using Bayes' 
theorem to update the probability distribution over the function space.  
This way, one obtains a data-informed \textit{posterior probability 
distribution} over the function space, adjusted with respect to the prior so as to be 
compatible with the observed data.  The posterior distribution may be used to
\textit{predict} (probabilistically) the value of the function at points where
the function has not yet been sampled.
The mean value of this GP prediction is our target 
interpolation/reconstruction for FDM and FVM.

Formally, a GP is a collection of random variables, any
finite collection of which has a joint Gaussian distribution
\cite{bishop2007pattern,rasmussen2005}.  
A GP is fully defined by two functions: 
\bit
\item a mean function $\bar{f}(\mathbf{x}) = \mathbb{E}[f(\bfx)]$ 
over $\mathbb{R}^N$, and 
\item a covariance function which is a symmetric, positive-definite 
integral kernel $K(\mathbf{x},\mathbf{y})$ over $\mathbb{R}^{N}\times\mathbb{R}^{N}$. 
\eit
Such functions $f$, drawn randomly from this distribution, are said to 
be sampled from a Gaussian Process with mean function $\bar{f}(\mathbf{x})$ 
and covariance function $K(\mathbf{x},\mathbf{y})$, and we write 
$f\sim\mathcal{GP}(\bar{f},K)$. %
As with the case of finite-dimensional Gaussian distributions, the significance 
of the covariance is
\begin{equation} 
K(\mathbf{x},\mathbf{y}) =
\mathbb{E}[
\left(f(\mathbf{x})-\bar{f}(\mathbf{x})\right)
\left(f(\mathbf{y})-\bar{f}(\mathbf{y})\right)
],
\label{eq:covariance_meaning}
\end{equation} 
where the averaging is over the GP distribution.

In standard statistical modeling practice, both $\bar{f}(\mathbf{x})$ and 
$K(\mathbf{x},\mathbf{y})$ are typically parametrized functions, with
parameters controlling the character (e.g. length scales, 
differentiability, oscillation strength) of ``likely'' functions. Given a 
GP, and given $N$ ``training'' points $\mathbf{x}_{i}$, $i=1,\ldots, N$ 
at which the function values $f(\mathbf{x}_{i})$ are known, we may 
calculate the likelihood $\mathcal{L}$ (the probability of $\mathbf{f}$ given the GP model) 
of the data vector 
$\mathbf{f}\equiv[f(\mathbf{x}_{1}),\ldots,f(\mathbf{x}_{N})]^{T}$ 
(e.g., $N$ many pointwise values of density $\rho$ at $\mathbf{x}_i$, $i=1, \ldots, N$)
by 
\beq
\mathcal{L}  \equiv  P(\mathbf{f})
 = (2\pi)^{-\frac{N}{2}}\det|\mathbf{K}|^{-\frac{1}{2}}
\exp\left[-\frac{1}{2}\left(\mathbf{f}-\bar{\mathbf{f}}\right)^{T}
\mathbf{K}^{-1}\left(\mathbf{f}-\bar{\mathbf{f}}\right)\right],
\label{eq:Like_1}
\eeq
where $\mathbf{K}=[{K}_{ij}]_{i,j=1,\dots, N}$ with $K_{ij}\equiv K(\mathbf{x}_{i},\mathbf{x}_{j})$.

Given the function samples 
$\mathbf{f}=[f(\mathbf{x}_{1}),\ldots,f(\mathbf{x}_{N})]^{T}$ 
obtained
at spatial points $\mathbf{x}_{i}$, $i=1,\ldots, N$, 
GP predictions aim to make a probabilistic statement about the value 
$f_{*}\equiv f(\mathbf{x}_{*})$
of the unknown function $f\sim\mathcal{GP}(\bar{f},K)$ at a new spatial point $\mathbf{x}_{*}$.
In other words, from a stochastic modeling view point,
we are interested in making a new prediction of GP for $f$ at any randomly chosen point $\mathbf{x}_{*}$.
This is particularly of interest to us from the perspectives of FDM and FVM, because
we can use GP to predict an unknown function value at cell interfaces (e.g., $u_{i\pm\frac{1}{2}}$ in 1D) 
where both FDM and FVM require estimates of flux functions.

We can accomplish this by utilizing the conditioning property of GP 
from the theory of Bayesian inference \cite{bishop2007pattern,rasmussen2005}.
We look at the augmented likelihood function $\mathcal{L}_{*}$ by considering
the joint distribution of the currently available training outputs, $\mathbf{f}$, and the new test output
$f_*$,
\beq
\mathcal{L}_{*}  \equiv  P\left(\mathbf{f},f_{*}\right)
  =  (2\pi)^{-\frac{N+1}{2}}\det|\mathbf{M}|^{-\frac{1}{2}}\exp\left[-\frac{1}{2}
  \left(\mathbf{g}-\bar{\mathbf{g}}\right)^{T}\mathbf{M}^{-1}\left(\mathbf{g}-
\bar{\mathbf{g}}\right)\right],
\label{eq:LikeStar_1}
\eeq
where $\mathbf{g}$ and $\bar{\mathbf{g}}$ are the $(N+1)$-dimensional
vectors whose components, in partitioned form, are
\begin{equation}
\mathbf{g}\equiv\left[f_{*},\mathbf{f}\right]^{T}, \;\;\; 
\bar{\mathbf{g}} \equiv\left[\bar{f}(\mathbf{x}_{*}),\bar{\mathbf{f}}\right]^{T},\label{eq:AugVec}
\end{equation}
and $\mathbf{M}$ is the $(N+1)\times(N+1)$ augmented covariance
matrix, given in partitioned form by
\begin{equation}
\mathbf{M}=\left(\begin{array}{cc}
k_{**} & \mathbf{k}_{*}^{T}\\
\mathbf{k}_{*} & \mathbf{K}
\end{array}\right).\label{eq:AugCov}
\end{equation}
In Eq.~(\ref{eq:AugCov}), we've defined a scalar $k_{**}$ and an
$N$-dimensional vector $\mathbf{k}_{*}=[\mbf{k}_{*,i}]_{i=1,\dots,N}$ given by
\beq
k_{**}  \equiv  K(\mathbf{x}_{*},\mathbf{x}_{*}), \;\;\; 
\mathbf{k}_{*,i} \equiv  K(\mathbf{x}_{*},\mathbf{x}_{i}).
\eeq

Using Bayes' Theorem, the conditioning property applied to
the joint Gaussian prior distribution on the observation $\mbf{f}$ 
yields the Gaussian posterior distribution of $f_*$ given $\mbf{f}$.
One may then straightforwardly derive \cite{bishop2007pattern,rasmussen2005}:

\begin{equation}
P(f_{*}|\mathbf{f})=\left(2\pi U^{2}\right)^{-\frac{1}{2}}
\exp\left[-\frac{\left(f_{*}-\bar{f_{*}}\right)^{2}}
{2U^{2}}\right],
\label{eq:Prediction}
\end{equation}
where the newly {\it{updated posterior mean function}} is
\begin{equation}
{\tilde {f}_{*}}\equiv
\bar{f}(\mathbf{x}_{*})+\mathbf{k}_{*}^{T} \mathbf{K}^{-1} \cdot \left(\mathbf{f}-\bar{\mathbf{f}}\right), 
\label{eq:PredMean}
\end{equation}
and the newly {\it{updated posterior covariance}} is
\begin{equation}
U^{2} \equiv k_{**}-\mathbf{k}_{*}^{T} \mathbf{K}^{-1} \cdot \mathbf{k}_{*}. 
\label{eq:PredVar}
\end{equation}

What has happened here is that the GP on the unknown function $f$, trained on the
data  $\mathbf{f}$, has
resulted in a Gaussian posterior probability distribution on the unknown function
value ${{f}(\bx_{*})}$ at a new desired location $\mathbf{x}_*$, 
with a mean
$\tilde {f}_*$ as given in Eq.~(\ref{eq:PredMean}), and with a variance as
given in Eq.~ (\ref{eq:PredVar}).

It is worthwhile to make an important observation at this point. 
Eq.~(\ref{eq:Like_1}) provides a clear conceptual difference 
in interpolations and reconstructions 
between the GP predictions and the RBF approximations.
In the GP {\it nonparametric} viewpoint we define a prior (or latent) probability distribution
{\it directly} over function spaces, hence there is no need to seek any fixed number of
parameters that may depend on the datasets under consideration.
This way, the stochastic properties of functions in the prior are set by a choice
of covariance kernel functions, by which the posterior predictions
in Eq. (\ref{eq:LikeStar_1}) follow.

On the contrary, RBF models in the context of interpolations and reconstructions 
follow the parametric modeling paradigm,
requiring to
solve a linear system $\bA \boldsymbol{\lambda} = \bF$ (see e.g., \cite{bigoni2016adaptive,Katz2009})
whose size is determined and {\it fixed}
by the number of desired $N$ interpolation points $\bff=[f(\bx_1), \dots, f(\bx_N)]^T$ and the associated 
coefficients (or parameters) $\lambda_j$ and $c_k$
of an RBF interpolating function
$s(\mbf{x})$,
\beq
s(\mbf{x}) = \sum_{j=1}^N \lambda_j \Phi(||\bx - \bx_j||, \epsilon_j ) + \sum_{k=1}^K c_k p_k(\bx).
\label{eq:RBF}
\eeq
The components of the solution vector $\boldsymbol{\lambda}$ are
$\lambda_j$ and $c_k$,
$\bA$ is the matrix whose components are set by the values of RBFs
$\Phi_{i,j}=\Phi(|| \bx_i - \bx_j||, \epsilon_j)$ at two points $\bx_i$ and $\bx_j$ and the shape parameter $\epsilon_j>0$,
$\bF$ is the vector with entries of $\bff$,
and the last term in Eq.~(\ref{eq:RBF}) is a polynomial (at most) $K$-th degree with
$k$-th standard basis polynomials $p_k$. 
Unlike GP, a probability distribution (if needed) is not directly imposed on $s(\bx)$,
rather, it can only be induced by a prior distribution defined on the coefficients.

Even though the two approaches resemble each other in terms of using ``kernels'' 
(i.e., RBFs and covariance kernel functions), they are fundamentally different from
these statistical viewpoints.

\subsection{High-order GP Interpolation for CFD}
\label{sec:gp_interpolation}
In this paper we are most interested in developing a high-order reconstruction method for FVM
in which the function samples $\mathbf{f}$ are given as volume-averaged data $\langle u_i \rangle$. However,
we first consider an interpolation method
which uses pointwise data $u_i$ as function samples $\mathbf{f}$. 
An algorithmic design for 
GP interpolation using pointwise data will provide a good mathematical foundation for FVM which
reconstructs pointwise values from volume-averaged data.

The mean ${\tilde {f}_{*}}$ of the
distribution given in Eq.~(\ref{eq:PredMean})
is 
our interpolation of the function $f$ at the point  $\mathbf{x}_{*} \in \mathbb{R}^D$, $D=1, 2, 3$,
where $f$ is any given fluid variable such as density, pressure, velocity fields, magnetic fields, etc.
For the purpose of exposition, let us use $q$ to denote one of such fluid variables (e.g., density $\rho = \rho(\bx,t^n)$).
Also, the mathematical descriptions of both interpolation and reconstruction will be considered in 1D hereafter.

We are interested in seeking a high-order interpolation of ${q}$
at $\mathbf{x}_*=x_*=x_{i\pm\frac{1}{2}}$ on a stencil $\cup_{k=i-p}^{i+r} I_{k}$,
\beq
{q}_{i\pm\frac{1}{2}}=\mcal{I}_{GP}(q_{i-p}, \dots, q_{i+r}), 
\eeq
where $\mcal{I}_{GP}(\cdot)$ 
is the GP interpolation given in Eq.~(\ref{eq:PredMean}).
We define
\beq
\label{eq:f_FDM}
\mathbf{f} = [q_{i-p}, \dots, q_{i+r}]^T.
\eeq
Furthermore, if we simply assume a constant mean $f_0$ for the data in Eq. (\ref{eq:f_FDM}) we get
\beq
\bar {\mathbf{f}} = f_0 \mathbf{1}_{r-p+1}, \;\;\; \bar{{f}}(x_{i\pm\frac{1}{2}}) = f_0,
\eeq
where $\mathbf{1}_{r-p+1}=[1, \dots, 1]^T$ is an $(r-p+1)$-dimensional one-vector.
In case of a zero mean $f_0=0$ the GP interpolation scheme simply becomes
\beq
\label{eq:gp_interpolation}
{q}_{i\pm\frac{1}{2}} = 
\mathbf{k}_{i\pm\frac{1}{2}}^{T} \mathbf{K}^{-1} \cdot \mathbf{f}.
\eeq
As shown in Eqs.~(\ref{eq:PredMean}) and (\ref{eq:gp_interpolation}), the 
interpolant $\tilde {f}_*=q_{i\pm\frac{1}{2}}$ is a simple linear combination of the observed data $\mathbf{f}$
and the covariance kernels $\mathbf{k_*}$ and $\mathbf{K}$,
anchored by one of its arguments to one of the 
data points, 
$x_*, x_{i-p}, \dots, x_{i+r}$.
The second term in Eq.~(\ref{eq:PredMean}) can also be cast as an 
inner product between a vector of weights 
$\mathbf{w}^T \equiv \mathbf{k}_{*}^T \mathbf{K}^{-1}$ and a vector of data 
residuals $(\mathbf{f}-\bar{\mathbf{f}})$.  

The weights $\mathbf{w}$ are independent 
of the data values $\mathbf{f}$; they depend only on the locations of the data points ${x}_i$ 
and on the desired interpolation point ${x}_{*}$.  
This is useful, because in hydrodynamic simulations  
the training point locations (viz. a GP
stencil) and the interpolation points (viz. cell interfaces) are often known in 
advance, as is the case with static Cartesian grid configurations.
In such cases, the weight vector $\mathbf{w}$ can be computed and stored in advance at an 
initialization step and remain constant throughout the simulation.
When an adaptive mesh refinement (AMR) configuration is considered,
$\mathbf{w}$ can be computed for all possible grid refinement levels and stored
{\it{a priori}} for later use. 
The GP interpolations then come at the cost of 
the remaining inexpensive inner product operation between $\mathbf{w}^T$ and $(\mathbf{f}-\bar{\mathbf{f}})$
in Eq. (\ref{eq:PredMean}), whose
operation count is linearly proportional to the number of points 
in the stencil. 
Typically in 1D, the size of the stencil is one for the first order Godunov (FOG) method
\cite{godunov1959difference,toro2009}; 
three for 2nd order piecewise linear methods (PLM) \cite{toro2009,van1979towards}; 
five for both the 3rd order piecewise parabolic method (PPM) \cite{colella1984piecewise};
and 5th order Weighted Essentially Non-oscillatory (WENO) method \cite{jiang1996efficient}.
Given such stencil sizes, the cost of the linear solves required by Eq.~(\ref{eq:PredMean}) are minor.
For instance, $\bK$ is a $5\times 5$ matrix when using a stencil of five grid points, 
namely, the 5-point GP stencil centered at $x_i$ with a radius of two-cell length $S_2=\cup_{k=i-2}^{i+2} I_k$.
In general, we define a stencil $S_R$ centered at $x_i$ with a radius $R$ cell-length,
\beq
\label{eq:GP_stencil_R}
S_R =\bigcup\limits_{k=i-R}^{i+R} I_{k}.
\eeq
Note that the interpolation point $x_*$ may be anywhere in the continuous
interval $S_R$.

Since the matrix $\mathbf{K}$ is symmetric positive-definite,
the inversion of $\mathbf{K}$
can be obtained efficiently by
Cholesky decomposition which is
about a factor 2 faster than the usual LU decomposition. The decomposition
is only needed once, at initialization time, for the calculation of the
vector of weights $\mathbf{w}$.

An important feature of GP interpolation is that it
naturally supports multidimensional stencil configurations. 
The reason for this is that there are many valid covariance functions
over $\mathbf{R}^D$ that are isotropic, and therefore do not bias
interpolations and reconstructions along any special direction.
The possibility of directionally-unbiased reconstruction over 
multidimensional stencils 
is a qualitative advantage of GP interpolation,
especially when designing 
high-order algorithms \cite{buchmuller2014improved,mccorquodale2011high,zhang2011order}.
Such high-order multidimensional GP methods will be studied in our future work. 

There is an additional piece of information beyond the point estimate 
$\tilde{f}_{*}$. We also have an uncertainty in the estimate, given by $U$ in 
Eq.~(\ref{eq:PredVar}).  This posterior uncertainty is of crucial importance 
in many GP modeling applications, but it is of limited interest 
for the purpose of this paper.
We will overlook posterior uncertainty in the current study, 
and focus on the posterior mean formula given in Eq.~(\ref{eq:PredMean}).

\subsection{High-order GP Reconstruction for CFD}
\label{sec:gp_reconstruction}
In FVM the fluid variables to be evolved are not pointwise values, but rather are
volume-averaged integral quantities, $\langle{q}_i\rangle=\frac{1}{\Delta \vol_i}\int_{\Delta \vol_i}q(\bx,t^n)d\vol$.
In FDM on the other hand, the main task is to find a high-order
approximation to the interface flux values $\hat{F}_{\iph}$, given the fact that
their integral quantities $F(q_i)=\frac{1}{\dx}\int_{x_\imh}^{x_\iph}\hat{F}(\xi)d\xi$ are known
via the analytic evaluations of the flux function $F$ at the given pointwise data set $u_i$.
The conservative FDM updates in 1D then readily proceed by solving
\beq
\label{eq:fdm}
\frac{dq_i}{dt} + \frac{1}{\dx} \left(\hat{F}_\iph - \hat{F}_\imh \right)= 0.
\eeq

In both cases, we see that there is a {\it{change in data types}} -- hence the name ``reconstruction'' -- 
between input (e.g., $\langle q_i\rangle$ for FVM; $F(q_i)$ for FDM) 
and output (e.g., $q(\bx,t^n)$ for FVM; $\hat{F}_\iph$ for FDM) 
pairs in such a way that
high-order approximations are applied to the integral quantities (inputs) and produce
corresponding pointwise values (outputs), with high accuracy.

The GP interpolation method outlined in Section \ref{sec:gp_interpolation}
should therefore be modified so that reconstruction may
account for such data type changes in both FDM and FVM.
Note that the integral averages over a grid cell 
constitute ``linear'' operations on a function $f(\mathbf{x})$. 
As with ordinary finite-dimensional multivariate Gaussian distributions,
where linear operations on Gaussian random variables result in new
Gaussian random variables with linearly transformed means and covariances,
a set of $N$ linear functionals operating on a GP-distributed function
$f$ has an $N$-dimensional Gaussian distribution with mean and covariance
that are linear functionals of the GP mean function and covariance
function.

Suppose for example in a FVM sense, we consider a GP reconstruction on a GP stencil $\cup_{k=i-p}^{i+r}I_k$.
First we define $r-p+1$ measures $dg_{k}(\mathbf{x})$,
$k=i-p,\ldots, i+r$, 
defining $r-p+1$ linear functionals,
\begin{equation}
G_{k}\equiv\int f(\mathbf{x}) \, dg_{k}(\mathbf{x}), \; k=i-p,\ldots, i+r.\label{eq:G_Functionals}
\end{equation}
For the sake of a reconstruction scheme on a mesh of control volumes, we choose 
the measures $dg_{k}(\mathbf{x})$ to be the
cell volume-average measures,

\begin{equation}
dg_{k}(\mathbf{x})=
\begin{cases}
\displaystyle d^3\mathbf{x}\times\prod_{d=x,y,z}\frac{ 1 } {\Delta^{(d)}} & \mbox{ if }  \mathbf{x}\in I_{k}, \\\\
0 & \mbox{ if }  \mathbf{x}\notin I_{k},
\end{cases}
\label{eq:g_cell}
\end{equation}
where $\Delta^{(d)}$ is the grid spacing in the $d$-direction, 
and the $k$-th cell $I_{k}$ is given by $I_k=\prod_{d=x,y,z} I_k\iter{d}$
with 1D cells $I_k\iter{d} = [x_k\iter{d}-\frac{\Delta\iter{d}}{2},x_k\iter{d}+\frac{\Delta\iter{d}}{2}]$ for
each $d$-direction. 
For the purpose of the current discussion, we will assume a locally-uniform
rectilinear grid of cubical cells $I_k$ of uniform size $\Delta=\Delta\iter{x}=\Delta\iter{y}=\Delta\iter{z}$.

Then, the vector $\mathbf{G}=[G_{i-p},\ldots,G_{i+r}]^{T}$ is normally
distributed with mean $\bar{\mathbf{G}}=[\bar{G}_{i-p},\ldots,\bar{G}_{i+r}]^{T}$
and covariance matrix $\mathbf{C}=\left[\mathbf{C}_{kh}\right]_{k,h = i-p, \dots, i+r}$, where
\beq
\bar{G}_{k}  =  
\mathbb{E}[G_{k}] =  
\int 
\mathbb{E}[f(\mathbf{x}) ] \, dg_{k}(\mathbf{x}) 
=  \int \bar{f}(\mathbf{x}) \, dg_{k}(\mathbf{x}) ,
\label{eq:G_mean}
\eeq
and
\begin{eqnarray}
\mathbf{C}_{kh} 
 & = &\mathbb{E}[\left(G_{k}-\bar{G}_{k}\right)\left(G_{h}-\bar{G}_{h}\right)]\nonumber \\
 & = & \int \mathbb{E} [\left(f(\mathbf{x})-\bar{f}(\mathbf{x})\right)\left(f(\mathbf{y})-\bar{f}(\mathbf{y})\right)] \, dg_{k}(\mathbf{x})\,dg_{h}(\mathbf{y})
\nonumber \\
 & = & \int  K(\mathbf{x},\mathbf{y}) \, dg_{k}(\mathbf{x}) \, dg_{h}(\mathbf{y}).
\label{eq:G_covariance}
\end{eqnarray}
Thus, we see that the GP distribution on the function $f$ leads to a multivariate 
Gaussian distribution on any $(r-p+1)$-dimensional vector $\mathbf{G}$ of linear 
functionals of $f$. 
This Gaussian distribution can be used for likelihood 
maximization in a manner completely analogous to 
the case of interpolation, where training is performed using pointwise data.

For reconstruction, it is also not hard to generalize Eq.~(\ref{eq:PredMean}).
We first define the $(r-p+1)$-dimensional \textsl{prediction vector} 
$\mbf{T}_*= \left[\mathbf{T}_{*,k}\right]_{k=i-p,\dots, i+r}$ at $\mbf{x}_*$
by 

\begin{equation}
\mathbf{T}_{*,k}\equiv{}
\int  K(\mathbf{x},\mathbf{x}_{*}) \, dg_{k}(\mathbf{x)}.
\label{eq:pred_vector}
\end{equation}
Then the pointwise function value ${f}(\bx_*)$ at the point $\mathbf{x}_{*}$, 
reconstructed from the volume-averaged data $\mbf{G}$, is given by
\begin{equation}
\tilde{f}_{*}=\bar{f}(\mathbf{x}_{*}) + 
\mathbf{T}_*^{T}\mathbf{C}^{-1}(\mathbf{G}-\bar{\mathbf{G}}).
\label{eq:vol_prediction}   
\end{equation}
Notice that the sample data $\bG$ now holds 
volume-averaged quantities on a stencil $\cup_{k=i-p}^{i+r}I_k$,
\beq
\bG=[\langle q_{i-p} \rangle, \dots, \langle q_{i+r} \rangle]^T.
\eeq
The $(r-p+1)$-dimensional vector
$\bT_{*,k}$ is the covariance between cell average quantities 
$\langle q_k\rangle, \, k=i-p, \dots, i+r$, and the pointwise value $q_*$ at $x_*$,
$\mathbf{C}$ is the $(r-p+1)\times (r-p+1)$ covariance kernel matrix between cell average values $\langle q_k\rangle$, and
$\bz=\mathbf{C}^{-1} \bT_{*,k}$ is the $(r-p+1)$-dimensional vector of weights for the data $(\bG-\bar\bG)$.
Eq. (\ref{eq:vol_prediction}) is a straightforward generalization of Eq.~(\ref{eq:PredMean}),
producing an integral version of it.

In a simple 1D case with a constant mean $f_0$ and $x_*=x_{i\pm\frac{1}{2}}$, the mean vector $\bar \bG$ is written as
\beq
\bar \bG= f_0{\mathbf{1}}_{r-p+1},
\eeq
and 
the GP reconstruction in Eq. (\ref{eq:vol_prediction}) becomes
\bea
\label{eq:gp_FVM}
\tilde{f}_{i\pm\frac{1}{2}} 
&=& f_0 + \mathbf{T}_{i\pm\frac{1}{2}}^{T}\mathbf{C}^{-1}(\mathbf{G}-f_0{\mathbf{1}}_{r-p+1}) \nonumber\\
&=& f_0 + \mathbf{z}^T(\mathbf{G}-f_0{\mathbf{1}}_{r-p+1}).
\eea

Analogous considerations arise in the RBF reconstruction methods for volume-averaged data.
In \cite{bigoni2016adaptive} the RBF function is integrated and is used for reconstruction, a procedure that corresponds closely
to integrating the kernel function in Eqs.~(\ref{eq:G_covariance}) and (\ref{eq:pred_vector}).

\subsection{The GP-SE Model}
\label{sec:gp_se}
One of the most widely-used kernels in GP modeling
is the ``Squared Exponential (SE)'' covariance kernel function 
\cite{bishop2007pattern,rasmussen2005}, which has the form
\begin{equation}
K\left(\mathbf{x},\mathbf{y}\right)\equiv
\Sigma^{2}\exp\left[-\frac{\left(\mathbf{x}-\mathbf{y}\right)^{2}}{2\ell^{2}}\right].
\label{eq:SE}
\end{equation}
This GP-SE model features three free parameters, $f_{0}$, $\Sigma^{2}$, and $\ell$.
Often, in Eq.~(\ref{eq:PredMean}) and Eq. (\ref{eq:vol_prediction}), 
a constant mean function 
$\bar{f}(\mathbf{x})=f_{0}\mbf{1}_N$ is adopted for simplicity, 
where $\mbf{1}_N=[1,\dots, 1]^T$ is an $N$-dimensional one-vector.
The constant $f_0$ can be analytically
determined from the data as part of the interpolation process by maximizing 
the likelihood function in Eq.~(\ref{eq:Like_1}). 
See Appendix \ref{sec:appendix_meanFunction} for more details.

The latter two, $\Sigma^{2}$ and $\ell$, are called ``hyperparameters'' 
which are the parameters that are built into the kernel.
The hyperparameter $\Sigma^{2}$ has no effect on the posterior mean function, 
so one can set $\Sigma^{2}=1$ for simplicity. 
On the other hand, the hyperparameter $\ell$ is the correlation length scale of the model.
It determines the length scale of variation preferred by the GP model.
Our GP predictions for interpolation/reconstruction, which necessarily agree with the observed values
of the function at the training points $\mbf{x}_k$, may ``wiggle'' on this scale between
training points. In this sense, $\ell$ is a ``rigidity'', controlling
the curvature scales of the prediction, and should correspond to the
physical length scales of the features GP is to resolve. 
Since we want
function interpolations/reconstructions that are smooth on the scale of the grid, we
certainly want $\ell>\Delta$, and would also prefer $\ell\geq R$.
As just mentioned, the choice of $\ell$ requires a
balance between the physical length scales in the problem and the grid
scale. This implies that different values of $\ell$ can be employed depending on
differing length scales on different regions of the computational domain.
We leave the investigation on best practice for determining such an optimal
$\ell$ within the context of FVM reconstructions 
to our forthcoming papers.
For the purpose of this paper we select a constant value of $\ell$ for each
simulation, which has a direct impact to the solution accuracy, as shown 
in Fig.~\ref{conv:length}.
The relationship often leads to the {\it trade-off principle} in that
there is a conflicting balance between the obtainable accuracy and the
numerical stability due to a large condition number of the kernel matrix
when a grid is highly refined.
Analogously in RBF theory, the trade-off principle also occurs
in terms of the shape parameters of the RBFs and it has been
investigated by several authors
\cite{Franke1982,fasshauer2007choosing,hardy1971multiquadric,rippa1999algorithm}.
More recent studies \cite{bigoni2016adaptive,guo2017rbf,jung2010recovery}
have focused on finding their optimal values of the shape parameters
in the combined ENO/WENO-RBF framework.
We discuss in detail our strategy of GP in Section \ref{sec:smooth-advection}.

The SE covariance function has two desirable 
properties \cite{bishop2007pattern,Cressie2015,rasmussen2005}.
First, it has the property of having a native space of $C^{\infty}$ functions,
which implies that the resulting interpolants converge
exponentially with stencil size, for data sampled from smooth functions.
Second, the SE kernel facilitates dimensional factorization, which is useful
in multidimensional cases.

Like all other GP kernels, 
SE suffers 
a notorious singularity issue where
the kernel is prone to yield nearly singular matrices when
the distance between any two points $\bx$ and $\by$
becomes progressively smaller (or equivalently, the grid becomes progressively refined in CFD). 
A practical and well-known fix for this problem is to add 
a ``nugget'' \cite{Cressie2015}, i.e., a small perturbation
$c_o \bI$ (where $c_0$ is a small positive constant, usually chosen
to be 10-100 times the floating-point machine precision) is added to $\bK$, where $\bI$ is the identity matrix.
Unfortunately, this trick does not resolve the issue in a desired way 
because it can result in less accurate data predictions in GP.
Our strategy to overcome the singularity issue in SE is discussed in Section \ref{sec:results}.

SE also performs poorly when the dataset contains discontinuities, such as shocks and contact discontinuities in a CFD context.
An intuitive resolution to this issue, from a statistical modeling perspective, would be to use
other types of non-smooth kernel functions instead such as
the Mat\'{e}rn class, exponential-types, rational quadratic functions, Wendland, etc.
\cite{bishop2007pattern,rasmussen2005,stein1999}.
They are known to be better-suited for discontinuous datasets than SE 
by relaxing the strong smoothness assumptions of SE. 
By construction, the latent (or prior) sample function spaces generated by these non-smooth kernels
possess finite-order of differentiability. As a result, we have seen that the
posterior GP predictions of interpolations and reconstructions using such kernels
turn out to be very noisy and rough, and their solutions
often suffer from reduced solution-accuracy and numerical instability.
A covariance kernel function such as the Gibbs covariance or the
neural-network covariance \cite{rasmussen2005} would be
more appropriate for resolving discontinuities. The possibility of using these GP kernels 
for high-order interpolations and reconstructions is under investigation and will be 
studied in our future work.
To deal with discontinuous solutions in this paper, 
we introduce yet another new approach to design
an improved GP reconstruction algorithm, 
termed as ``GP-WENO'', which is described in Section \ref{sec:gp-weno}.
This new approach formulates a new set of non-polynomial, GP-based smoothness indicators
for discontinuous flows.

\begin{figure}[tppb!] 
\centering
 \includegraphics[width=.4 \textwidth]{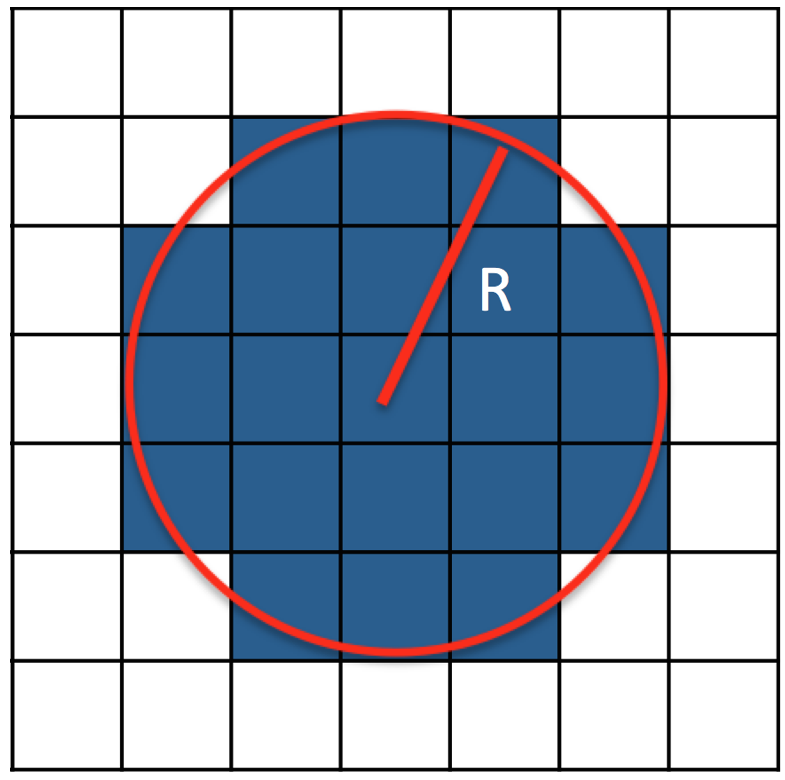}
\caption{{\footnotesize An example of a 2D GP stencil with a radius $R$. It consists of a collection grid cells $I_{i_k}$ 
(blue cells) that are
included in a multidimensional blocky sphere of a radius $R$ from the local cell $I_{i_0}$,
in which the GP interpolation/reconstruction takes place. 
}}

 \label{fig:gp_stencil}
\end{figure}

Although we are interested in describing one-dimensional formulations of GP in this paper,
let's briefly discuss a 2D case that best illustrates a general strategy of 
GP to compute a single interpolation and reconstruction procedure.
Consider a list of $N$ cells $I_{i}$, $i=1, \dots, N$ in 2D. 
The most natural way to construct the list of cells (i.e., the stencil)
over which the GP interpolation/reconstruction will take place
is to pick a radius\footnote[$\dagger$]{In the current study
$R$ is an integer multiple of $\Delta$ for simplicity, which needs not be the case in general.} $R$, 
and add to the list those cells $I_{i_k}$ whose cell centers $\mbf{x}_{i_k}$
are within the distance $R$ from a local cell $I_{i_0}$ under consideration.
The result is a ``blocky sphere'' that ensures isotropy of the
interpolation. See Fig.~\ref{fig:gp_stencil}. We can adjust $R$ to
regulate $N$,
thus making $R$ a performance/accuracy
tradeoff tuning parameter. 

An important practical feature of the SE covariance function is that it provides
its dimensional factorization analytically.
Therefore, the volume averages in 
Eqs.~(\ref{eq:G_mean}) and (\ref{eq:G_covariance}) simplify to iterated 
integrals, and in fact they can be expressed {\it{analytically}} in terms of a 
pre-computed list of error functions of arguments proportional to 
one-dimensional cell center differences. Eqs.
(\ref{eq:G_covariance}) and (\ref{eq:pred_vector}) become
\begin{align}
  \label{eq:SE-cov}
  \mathbf{C}_{kh} =  \sqrt{\pi}\left ( \ \frac{\ell}{\Delta} \right
                       )^2 & \left \{ \left (
    \frac{\Delta_{kh}+1}{\sqrt{2}\ell/\Delta}\erf{\frac{\Delta_{kh}+1}{\sqrt{2}\ell/\Delta}}
  +
  \frac{\Delta_{kh}-1}{\sqrt{2}\ell/\Delta}\erf{\frac{\Delta_{kh}-1}{\sqrt{2}\ell/\Delta}}
                       \right ) \right .
   \nonumber \\
  +  \frac{1}{\sqrt{\pi}} & \left . \left (
              \expo{-\frac{(\Delta_{kh}+1)^2}{2(\ell/\Delta)^2}} + \expo{-\frac{(\Delta_{kh}-1)^2}{2(\ell/\Delta)^2}} \right
              ) \right . \nonumber \\ 
              -2 & \left . \left (
                 \frac{\Delta_{kh}}{\sqrt{2}\ell/\Delta}\erf{\frac{\Delta_{kh}}{\sqrt{2}\ell/\Delta}}
                 + \frac{1}{\sqrt{\pi}}\expo{-\frac{\Delta_{kh}^2}{2(\ell/\Delta)^2}}\right ) \right \},
\end{align}
and 
\begin{equation}
  \label{eq:SE-pred}
  \bT_{*,k} = \sqrt{\frac{\pi}{2}}\frac{\ell}{\Delta} \left \{
     \erf{\frac{\Delta_{k*}+1/2}{\sqrt{2}\ell/\Delta}}
     - \erf{\frac{\Delta_{k*}-1/2}{\sqrt{2}\ell/\Delta}}
     \right \}, 
\end{equation}
using the SE kernel, where $\Delta_{kh}=(x_k-x_h)/\Delta$. Other choices of covariance kernel functions
are also available \cite{stein1999,Cressie2015}, while 
such choices would require numerical approximations for 
Eqs.~(\ref{eq:pred_vector}) and (\ref{eq:G_covariance}), rather than a possible analytical form
as in the SE case.

\subsection{GP-WENO: New GP-based Smoothness Indicators for Non-smooth Flows}
\label{sec:gp-weno}

For smooth flows 
the GP linear prediction
in Eq.~(\ref{eq:vol_prediction}) 
with the SE covariance kernel 
Eq.~(\ref{eq:SE}) 
furnishes a high-order GP reconstruction algorithm
without any extra controls on numerical stability.
However, for non-smooth flows the unmodified
GP-SE reconstruction suffers from unphysical oscillations that originate at discontinuities such as shocks.
To handle flows with
discontinuities a hybrid method can be implemented using a shock
detector (see \cite{balsara1999}) to switch to a lower order
piecewise linear method from the GP reconstruction when there is a shocked cell in the
stencil. We have implemented such a hybrid method 
and found that it works well in general for problems with shocks,
such as the Sod shock tube problem. 
However, we noticed that it fails to capture features in 
flow regions that contain a transition from smooth flow to a shock, 
producing unphysical oscillations there.

To ultimately resolve such issues with non-smooth flows we adopt the principal idea of 
employing the nonlinear weights in the
Weighted Essentially Non-oscillatory (WENO) methods
\cite{jiang1996efficient}, by which we adaptively change the size of the reconstruction
stencil to avoid interpolating through a discontinuity, while
retaining high-order properties in smooth flow regions. A traditional WENO takes the
weighted combination of candidate stencils based on the local
smoothness of the individual sub-stencils. The weights are chosen so
that they are optimal in smooth regions, in the sense that they are
equivalent to an approximation using the global stencil that is the
union of the candidate sub-stencils. 

For a stencil $S_R$ of a radius $R$, with $2R+1$ points centered at
the cell $I_i$, we consider $R+1$
candidate sub-stencils $S_m \subset S_R$, each with $R+1$ points. 
Similar to the traditional WENO schemes,
we subdivide $S_R$ given as
\beq
  \label{eq:stencils}
  S_R  = \bigcup\limits_{m=i-R}^{i+R}I_m 
\eeq
into $R+1$ sub-stencils $S_m$,
\beq
  \label{eq:stencil2}
  S_m  = \{I_{i-R+m-1}, \ldots, I_i, \ldots, I_{i+m-1} \}, \;\;\; m=1,\ldots,R+1.
\eeq
They satisfy
\beq
S_R=\bigcup\limits_{m=1}^{R+1}S_m, \mbox{ and } I_i=\bigcap\limits_{m=1}^{R+1}S_m.
\eeq

\begin{figure}[htbp!]
  \centering
    \label{conv:gp_beta}
  \includegraphics[width=0.8\textwidth]{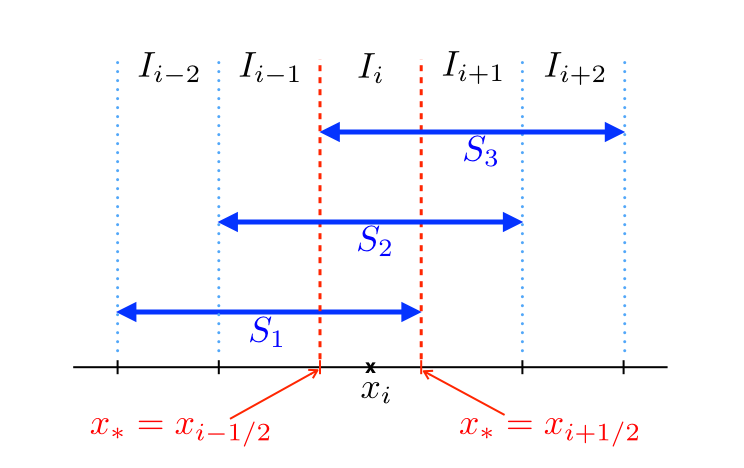}
  \caption{A schematic example of the 5-point GP stencil with $R=2$ centered at $x_i$. 
  Two grid locations indicated as $x_*$ are the target locations where we seek high-order GP 
  approximations of pointwise value (e.g., $\rho_{*}$)
  using the given cell-centered volume average data nearby (e.g., $ \langle {\rho}_s\rangle$, $s=i-2, \dots, i+2$).}
\end{figure}

Using Eq.~(\ref{eq:vol_prediction}) on each sub-stencil $S_m$, the GP approximation 
for a pointwise value
at a cell location $x_*$  
from the $m$-th sub-stencil $S_m$
is
\begin{equation}
  \label{eq:Sm_prediction}
  \tilde{f}_*^{m} = f_0 + \bfz^T_m (\bfG_m - f_0\mathbf{1}_{R+1}).
\end{equation}
On the other hand, from the global $2R+1$ stencil $S_R$ we also have a pointwise GP approximation at the same location
\beq
  \label{eq:SR_prediction}
  \tilde{f}_* = f_0 + \bfz^T (\bfG - f_0\mathbf{1}_{2R+1}).
\eeq
Here, the cell location ``$*$" is where we want to obtain a desired high-order reconstruction from GP,
e.g., the cell interface $x_*=x_{i\pm 1/2}$ is of particularly of interest to the Riemann problems in our case.
Further, $f_0$ is a constant mean function that is the same over all the
candidate sub-stencils, $\bfz^T_m=\bfT_{*,m}\bfC^{-1}_m$ is the $(R+1)$-dimensional vector of
weights, $\bfG_m$ is the $(R+1)$-dimensional vector of volume averaged data on $S_m$ (e.g., the volume averaged densities), 
and $\mathbf{1}_{R+1}=[1, \ldots, 1]^T$ is the $(R+1)$-dimensional one-vector. 
Likewise, $\bfz^T=\bfT_* \bfC^{-1}$, $\bfG$, and 
$\mathbf{1}_{2R+1}$ are $(2R+1)$-dimensional vectors defined in the same way.

We now take the weighted combination of these GP
approximations as our final reconstructed value
\begin{equation}
  \label{eq:weighted_sum}
  \tilde{f}_* = \sum_{m=1}^{R+1}\omega_m\tilde{f}_*^m.
\end{equation}
As in the traditional WENO approach, the weights $\omega_m$ should reduce to some
optimal weights $\gamma_m$ in smooth regions 
such that the approximation in
Eq.~(\ref{eq:weighted_sum}) gives the GP approximation over the global
$2R+1$ point stencil. The $\gamma_m$'s then should satisfy
\begin{equation}
  \label{eq:optimal_weights}
  \bz^T(\bG - f_0\mathbf{1}_{2R+1}) = \sum_{m=1}^{R+1}\gamma_m
  \bz^T_m(\bG_m - f_0\mathbf{1}_{R+1}), 
\end{equation}
where $\boldsymbol{\gamma} = [\gamma_1, \dots, \gamma_{R+1}]^T$ is given by the solution to the $(R+1) \times (2R+2)$
overdetermined system
\begin{equation}
  \label{eq:gamm_solns}
  \left (
  \begin{array}{c}
    \bM \\
    \sum_{m=1}^{R+1} f_0 \bz^T_{m} \cdot \mathbf{1}_{R+1} 
  \end{array}
  \right )
  \boldsymbol{{\gamma}} = 
  \left (
    \begin{array}{c}
      \bz \\
      f_0 \bz \cdot 
      \mathbf{1}_{2R+1}
    \end{array}
  \right ).
\end{equation}
The columns of the matrix $\bM$ are given by
\begin{equation}
  \label{eq:M-mat}
  \bM_{km} = \left \{
    \begin{array}{cl}
      \bz^T_{k,m}& \text{if } I_k \in S_m, \\
      0          & \text{otherwise}.
    \end{array}
  \right . 
\end{equation}
The optimal weights $\gamma_m$ then depend on the choice of kernel
function, but as with the vectors of GP weights, $\bz$ and $\bz_m$, $\gamma_m$
need only be computed once. We take the $\gamma_m$'s as the least
squares solution to the overdetermined system in
Eq.~(\ref{eq:gamm_solns}), which can be determined numerically. 

It remains to describe the non-linear weights $\omega_m$ in
Eq.~(\ref{eq:weighted_sum}). Again, these should reduce to the optimal weights
in smooth regions, but more importantly, they need to serve as an indicator
of the quality of data on the sub-stencil $S_m$. We adopt the same weights
as in the WENO-JS methods \cite{jiang1996efficient}:
\begin{equation}
\label{eq:WENO-JS_omega}
\omega_m^{} = \frac{\tilde{\omega}_m^{}}{ \sum_{s}\tilde{\omega}_s^{}}, \mbox{ where }
\tilde{\omega}_m^{} = \frac{\gamma_m^{}}{(\epsilon + \beta_m)^p},
\end{equation}
where we set $p=1$ and $\epsilon=10^{-36}$ for our test results.
These weights are based on the so-called smoothness indicators
$\beta_m$, for each stencil. In modern WENO schemes the smoothness
indicators are calculated as the $L_2$-norm of the reconstructed
polynomial on each ENO stencil $S_m$. We wish to construct a new class
of smoothness indicators that fits within the GP framework.

To begin with, we give a brief description on the eigenfunction analysis of covariance kernels.
Gaussian process regression, as described in
Section~\ref{sec:Gaussian-processes}, can be equivalently viewed as
Bayesian linear regression using a possibly infinite number of basis
functions. One choice of basis functions is the eigenfunctions,
$\phi_i(\cdot)$, $i=1, \dots, n$, of the covariance kernel function, $k$. 
By definition each eigenfunction $\phi_i$ satisfies
\begin{equation}
  \label{eq:Efcn}
  \int k(x,x') \phi_i(x) d\mu(x) = \lambda_i \phi_i(x').
\end{equation}
The eigenpair $(\phi_i, \lambda_i)$ consists of the eigenfunction and the eigenvalue
of the covariance kernel function $k$ with respect to the measure $\mu$
\cite{bishop2007pattern,rasmussen2005}. 

Of interest is when there is 
a non-negative function $p(x)$ called the {\it density} of the measure $\mu$
so that the
measure may be written as $d\mu(x) = p(x)dx$. We can approximate the integral in
Eq.~(\ref{eq:Efcn}) by choosing $n$ finite sample points, $x_l$, $l=1, \dots, n$, 
from $p(x)$ on some stencil, which gives
\begin{equation}
  \label{eq:apprx_eig}
  \lambda_i\phi_i(x') \simeq \frac{1}{n}\sum_{l=1}^{n}k(x_l,x')\phi_i(x_l).
\end{equation}
Plugging in $x'=x_l$ into Eq. (\ref{eq:apprx_eig}) for $l=1, \dots, n$
gives the matrix eigenproblem
\begin{equation}
  \label{eq:matrix_eigen}
  \bK \bvv_i = n\lambda_i\bvv_i,
\end{equation}
where $\bK$ is an $n\times n$ Gram matrix with entries
$\bK_{ij}=k(x_i,x_j)$, and $n\lambda_i$ is the $i$-th matrix eigenvalue
associated with the normalized $n$-dimensional eigenvector $\bvv_i$ satisfying 
$\bvv_i^T \bvv_j = \delta_{ij}$. We see that
$\phi_i(x_j) \sim \sqrt{n} \be_j \cdot \bvv_i$ where $\be_j$ is the $j$-th unit vector, and we get the $\sqrt{n}$
factor from differing normalizations of the eigenvector and eigenfunction.
This eigendecomposition of the covariance matrix, when viewed as a principal
component analysis (PCA) \cite{bishop2007pattern,rasmussen2005}, can be thought of as decomposing the
covariance into the various variational modes that can be represented
on the $n$ sample points. For more discussion of eigenfunction
analysis of covariance kernels we refer readers to Section 4.3 of
\cite{rasmussen2005}; Section 12.3 of \cite{bishop2007pattern}.

A set of $n$ samples of a function $f(x)$, denoted as
\beq
\bff = [f(x_1), f(x_2), \dots, f(x_n)]^T,
\eeq
evaluated at the points $x_l, \, l=1, \dots, n$
as in Eq.~(\ref{eq:apprx_eig}), can be expanded in the eigenbasis in
(\ref{eq:matrix_eigen}) as 
\beq
\mathbf{f} = \sum_{i=1}^{n}\alpha_i \bvv_i.
\eeq

We can then define a norm in the Hilbert space of functions
using the inner product of the function in the
reproducing kernel Hilbert space (RKHS) 
defined by the kernel function
$k(x,x')$ \cite{rasmussen2005},
\beq
\label{eq:Hnorm}
\| f\|^2_\mathcal{H} = <f,f>_\mathcal{H}= \sum_{i=1}^{n} \frac{\alpha_i^2}{\lambda_i}.
\eeq 
In this expression for the norm $\|f\|^2_\mathcal{H}$, the 
relative weight of a given mode in the data $\mathbf{f}$ is inversely
weighted by the eigenvalue $\lambda_i$. 
The SE kernel has a native
Hilbert space of $C^\infty$ functions, and assigns relatively small eigenvalues
to eigenfunctions that vary rapidly on short length scales.  The quantity $\|f\|^2_\mathcal{H}$ thus contains
large additive terms when the data embodies a discontinuity, while
data corresponding to
smooth regions produces smaller additive terms. This suggests that we may
use $\|f\|^2_\mathcal{H}$ to supply the basis for a GP-based 
smoothness indicator in the non-linear weights in Eq. (\ref{eq:WENO-JS_omega}). 

We now use the above discussion of the eigendecomposition to focus on
the practical implementation of our new GP-based smoothness indicators.
It remains to specify how to obtain the $\alpha_i$'s from the volume
averaged data $\bfG_m$ on the sub-stencil $S_m$. 
Using the notation `$m$' to denote the objects restricted on each $S_m$,
the eigenproblem in Eq. (\ref{eq:matrix_eigen}) on $S_m$ can be written as
\begin{equation}
  \label{eq:matrix_eigen_Sm}
  \bK^{m} \bvv_i^{m} = \lambda_i^{m}\bvv_i^{m}, \, i = 1, \dots, R+1,
\end{equation}
where $\bK^{m}$ is the $(R+1)\times(R+1)$ sub-matrix.
Note that here we are using the {\it pointwise} covariance kernel $\bK^{m}$
from Eq.~(\ref{eq:SE}) rather than the {\it integrated} covariance kernel $\bC^{(m)}$ from Eq.~(\ref{eq:G_covariance}).
In this way we obtain the eigenpair $(\lambda_i^{m}, \buu_i^{m})$ that is consistent with
providing $\beta_m$ which is to measure
smoothness of {\it pointwise} values $\tilde{f}^m_* = \tilde{f}^{m}(x_*) = \tilde{f}^{m}(x_{i\pm \frac{1}{2}})$ 
in Eq.~(\ref{eq:weighted_sum}).

Consider a $(R+1)$-dimensional
cell-centered pointwise values
\beq
\mathbf{f}^{m}=[f^m(x_1), \dots, f^m(x_{R+1})]^T.
\eeq
We can express $\mathbf{f}^m$ in the basis $\bvv_i^{m}$ as
\beq
\label{eq:fm_span}
\mathbf{f}^m = \sum_{i=1}^{R+1} \alpha_i^m \bvv_i^{m}, \, \mbox{ for each } m=1, \dots, R+1.
\eeq
Each element $f^m(x_l)$, $l=1, \dots, R+1$,  can be reconstructed from $\bfG_m$, using a
zero mean $f_0=0$ in Eq. (\ref{eq:gp_FVM}), as
\begin{equation}
  \label{eq:rec_fm}
  f^{m}(x_l) = \bfz^T_m(x_l)\cdot\bfG_m,
\end{equation}
where $(R+1)$-dimensional vector $\bfz^T_m(x_l) = \bfT_{l,k}^m(\bfC^m)^{-1}$, $k=1, \dots, R+1$,
and
\beq
\bG_m=[\langle q_1\rangle, \dots, \langle q_{R+1}\rangle]^T.
\eeq
Here  $\bfT_{l,k}^m$ is the
prediction vector (Eq.~(\ref{eq:SE-pred})) over the sub-stencil $S_m$
evaluating the covariance between the cell average values $\langle q_k\rangle$, $k=1, \dots, R+1$,
and the pointwise value $q_l$ evaluated at each $x_l \in S_m$.

Obtaining $\alpha_i^m$ is straightforward from Eq.~(\ref{eq:fm_span}). Using $(\bvv_j^m)^T\bvv_i^m=\delta_{ji}$,
\beq
\label{eq:alphas_1}
\alpha_i^m = (\bvv_i^m)^T\cdot \mathbf{f}^m.
\eeq
Finally for each $m$ we have our new GP-based smoothness indicators for the
data on the sub-stencil $S_m$ given by
\begin{equation}
  \label{eq:betas}
  \beta_m = \sum_{i=1}^{R+1}\frac{\left ( \alpha_i^{m} \right )^2}{\lambda^m_i}.
\end{equation}

The computational efficiency of Eqs. (\ref{eq:rec_fm}) and (\ref{eq:alphas_1}) can be
improved by introducing 
vectors $\bfP^{m}_i$ defined as
\begin{equation}
  \label{eq:Pvec}
  \bfP^{m}_i = (\bvv^m_i)^T \cdot \bfZ^T_m,
\end{equation}
where the $l$-th column of the matrix $\bfZ_m$ is given by
$\bfz_m(x_l)$, $l=1, \dots, R+1$.
In this way we precompute $\bfP^{m}_i$ which can directly act on the volume average quantities of $\bG_m$, 
by which the $\alpha_i^m$'s are given by
\begin{equation}
  \label{eq:alphas_2}
  \alpha^{m}_i = \bfP^{m}_i \cdot \bfG_m.
\end{equation}
To see the computational efficiency in comparison, in Appendix \ref{sec:appendix_operationCounts}
we provide operation counts of the two approaches  
referred to as ``{Option 1}'' with solving Eqs. (\ref{eq:rec_fm}) and (\ref{eq:alphas_1}), 
and ``{Option 2}'' with solving Eqs. (\ref{eq:Pvec}) and (\ref{eq:alphas_2}).
{Option 2} was used for the results shown in Section \ref{sec:results}.

It is noteworthy that there is an alternative interpretation of Eq. (\ref{eq:betas}) 
available in terms of the log likelihood function.
Notice that Eq. (\ref{eq:betas}) can be written as a quadratic form
\beq
\label{eq:log_likelihood}
(\mathbf{f}^m)^T (\bK^{m})^{-1} \mathbf{f}^m,
\eeq
which is again equivalent to
\beq
-2\ln{P(\mathbf{f}^m)} + C,
\eeq
assuming $f_0=0$  in the relation in Eq.~(\ref{eq:Like_1}) and
$C=\ln(\det{|\mathbf{K}|}^{-1})+N\ln(2\pi)$.
The analogy between Eq.~(\ref{eq:betas}) and Eq.~(\ref{eq:log_likelihood}) 
can be proved easily if we realize that the matrix $\bK^{m}$ can be expressed as
\beq
\bK^{m} = \sum_{i=1}^{R+1} \lambda_i^m \bvv_i^{m} (\bvv_i^{m})^T,
\eeq
and hence
\beq
\label{eq:K_inverse}
(\bK^{m})^{-1} = \sum_{i=1}^{R+1} \frac{1}{\lambda_i^m} \bvv_i^{m} (\bvv_i^{m})^T.
\eeq
The result follows easily by plugging 
Eqs.~(\ref{eq:fm_span}) and (\ref{eq:K_inverse}) into Eq.~(\ref{eq:log_likelihood}) to get
Eq.~(\ref{eq:betas}).
So the statement $\beta_m$ is large in this sense is analogous to the statement that
the likelihood of $\mathbf{f}^m$ is small.
The statistical interpretation is that short length-scale variability in 
$\mathbf{f}^m$ make its data unlikely according to the smoothness model represented by $\bK^{m}$.

It was observed by Jiang and Shu \cite{jiang1996efficient} that it is
critically important to have appropriate smoothness indicators so that
in smooth regions the non-linear weights reproduce the optimal linear
weights to the design accuracy in order to maintain high order of
accuracy in smooth solutions. Typically this is shown by Taylor
expansion of smoothness indicators in Eq.~(\ref{eq:WENO-JS_omega}). Here
we propose to calculate the eigendecomposition in
Eq.~(\ref{eq:matrix_eigen}) numerically, and therefore we don't provide a
mathematical proof for the accuracy of our new smoothness
indicators. Nonetheless we observe the expected high order convergence
expected from the polynomial WENO reconstructions on the equivalent
stencils in numerical experiments provided in
Section~\ref{sec:smooth-advection}.

\section{Steps in GP-SE Algorithm for 1D FVM }
\label{sec:gp_1d_steps}
Before we present the numerical results of the GP reconstruction algorithms detailed in the above,
we give a quick summary on the step-by-step procedure for 1D simulations with FVM.
The 1D algorithm of GP outlined above can be summarized as below:

\renewcommand{\labelenumi}{Step \arabic{enumi}}
\renewcommand{\labelenumii}{Step (\alph{enumii})}
\begin{enumerate}
\item \textbf{Pre-Simulation:} The following steps are carried out
  before starting a simulation, and any calculations therein need only
  be performed {\it once, stored, and used} throughout the actual simulation.

  \begin{enumerate}
  \item \textbf{Configure computational grid:} Determine a GP stencil
    radius $R$ as well as choose the size of the hyperparameter
    $\ell$. This determines the SE kernel function in
    Eq.~(\ref{eq:SE}) as well as the global and candidate stencils in
    Eqs.~(\ref{eq:stencils}) \& (\ref{eq:stencil2}).

  \item \textbf{Compute GP weights:} Compute the covariance
    matrices, $\bC$ and $\bC_m$ (Eq.~(\ref{eq:G_covariance})) on the stencils $S_R$ and
    each of $S_m$ as well as the prediction vectors, $\bT_*$
    (Eq.~(\ref{eq:pred_vector})). The GP weight
    vectors, $\bz^T=\bT_*\bC^{-1}$, on each of the stencils can then be stored for use in the
    GP reconstruction. The columns of the matrices $\bfZ_m$,
    $\bfz_m(x_l)$'s, should be computed here for use in step (d). It is crucial in this step as well as in Step
    (c) to use the appropriate floating point precision to prevent the
    covariance matrix $\bC$ from being numerically singular. This is
    discussed in more detail in Section~\ref{sec:smooth-advection}. We
    find that double precision is suitable up to condition numbers
    $\kappa\sim 10^8$, whereas quadruple precision allows up to
    $\kappa\sim 10^{18}$. 
    The standard double precision is used except for Step (b) and Step (c).

  \item \textbf{Compute linear weights:} Use the GP weight vectors to
    calculate and store the optimal linear weights according to Eq.~(\ref{eq:gamm_solns}).

  \item \textbf{Compute kernel eigensystem:} The eigensystem for the
    covariance matrices used in GP-WENO are calculated
    using Eq.~(\ref{eq:matrix_eigen}). The matrices, $\bC_m$,
    are the same on each of the candidate stencils in the GP-WENO
    scheme presented here, so only one eigensystem needs to be
    determined. This eigensystem is then used to calculate and store the vectors
    $\bfP^{(m)}_i$ in Eq.~(\ref{eq:Pvec}) for use in determining the
    smoothness indicators in the reconstruction Step 2.
  \end{enumerate}

  At this point, before beginning the simulation, if quadruple
  precision was used in Step (b) and Step (c) the GP weights and linear
  weights can be truncated to double precision for use in the actual
  reconstruction step.

\item \textbf{Reconstruction:} Start a simulation. 
  Choose $f_0$ according to Eq.~(\ref{eq:f_0_reconstruction})
  or simply set to zero.
  The simplest choice $f_0 = 0$ gives good results in practice
  and yields $\bar{\mathbf{G}} = 0$ in Eq. (\ref{eq:vol_prediction}).
  At each cell $x_i$, calculate the updated posterior mean function
  $\tilde{f}_*$ in Eq. (\ref{eq:vol_prediction}) as a high-order GP reconstructor to compute 
  high-order pointwise Riemann state values at $x_*=x_{i\pm 1/2}$
  using each of the candidate stencils. The
  smoothness indicators (Eq.~(\ref{eq:betas})) are calculated using
  the eigensystem from Step (d) and, in conjunction with the linear weights
  from Step (c), form the non-linear weights
  (Eq.~(\ref{eq:WENO-JS_omega})). Then take the convex combination
  according to Eq.~(\ref{eq:weighted_sum}).

  \item \textbf{Calculate fluxes:} Solve Riemann problems at cell interfaces using 
the high-order GP Riemann states in Step 2 as inputs.

  \item \textbf{Temporal update:} Update the volume-averaged solutions $\langle q_i\rangle$ 
from $t^n$ to $t^{n+1}$ using the Godunov fluxes from Step 3.
\end{enumerate}

\section{Numerical Results}
\label{sec:results}

Here we present numerical results using 
the GP-WENO reconstructions with stencil radii $R=1,2,3$ (denoted as GP-R1, GP-R2, GP-R3 respectively), 
described in Section~\ref{sec:gp-weno}, applied to the 1D compressible Euler
equations and the 1D equations of ideal
magnetohydrodynamics (MHD). 
The GP-WENO method will be the default reconstruction scheme hereafter.

We compare the solutions of GP-WENO with the fifth-order WENO (referred to as WENO-JS in what follows)
FVM \cite{jiang1996efficient,shu2009high}, using the same non-linear weights in Eq. (\ref{eq:WENO-JS_omega}).
The only difference between using WENO-JS and GP-R2
is the use of polynomial based reconstruction and smoothness indicators \cite{shu2009high} for WENO-JS, 
and the use of Gaussian
process regression with the new GP-based smoothness indicators for GP. A fourth-order TVD
Runge-Kutta method \cite{shu1988tvd} for temporal updating, and the HLLC
\cite{li2005hllc,toro1994restoration} or Roe
\cite{roe1981approximate} Riemann solvers are used
throughout. 

\subsection{Performance Comparison}

We first present a performance comparison for the proposed GP scheme
and the WENO-JS scheme summarized in Table~\ref{tab:performance}. We see
that the cost of
the GP-R2 and WENO-JS methods are very nearly the same, operating on
the same stencils. The minor
difference can be attributed to the new GP based smoothness
indicators. It can be seen in Eq.~(\ref{eq:betas}) the GP smoothness
indicators can be written as the sum of $R+1$ perfect squares, which
is three terms for each $\beta_m$ for GP-R2, meanwhile the smoothness indicators for WENO-JS
contain only two terms (i.e., the first and second derivatives of the second-degree ENO polynomials $p_m(x)$). 
The additional cost is offset by a significant
advantage in smooth regions near discontinuities over the WENO-JS
scheme, which is discussed in Section~\ref{sec:shu-osher-shock}.

\begin{table}[h!]
  \centering
  \begin{tabular}{||c|c||}
    \hline
    Scheme & Speedup \\ \hline
    GP-R1   & 0.8 \\
    GP-R2   & 1.0 \\
    GP-R3   & 1.4 \\
    WENO-JS  & 0.9 \\ \hline 
  \end{tabular}
  \caption{Shown is the relative time to solution for the four methods
    considered, all normalized to the GP-R2 time.}
  \label{tab:performance}
\end{table}

\subsection{1D Smooth Advection}
\label{sec:smooth-advection}

\begin{figure}[htbp!]
  \centering
  \subfigure[][]{
    \label{conv:plot}
  \includegraphics[width=0.8\textwidth]{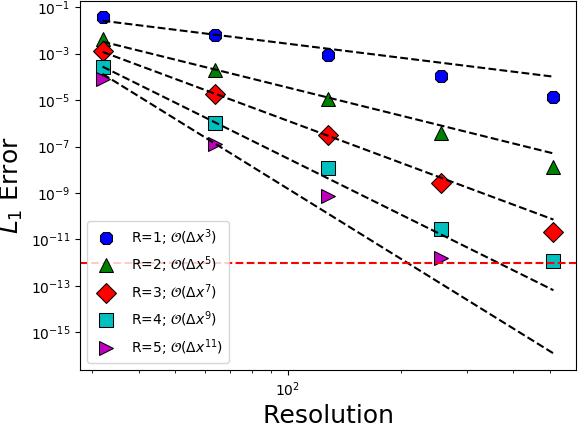}}
  \subfigure[][]{
    \label{conv:length}
  \includegraphics[width=0.8\textwidth]{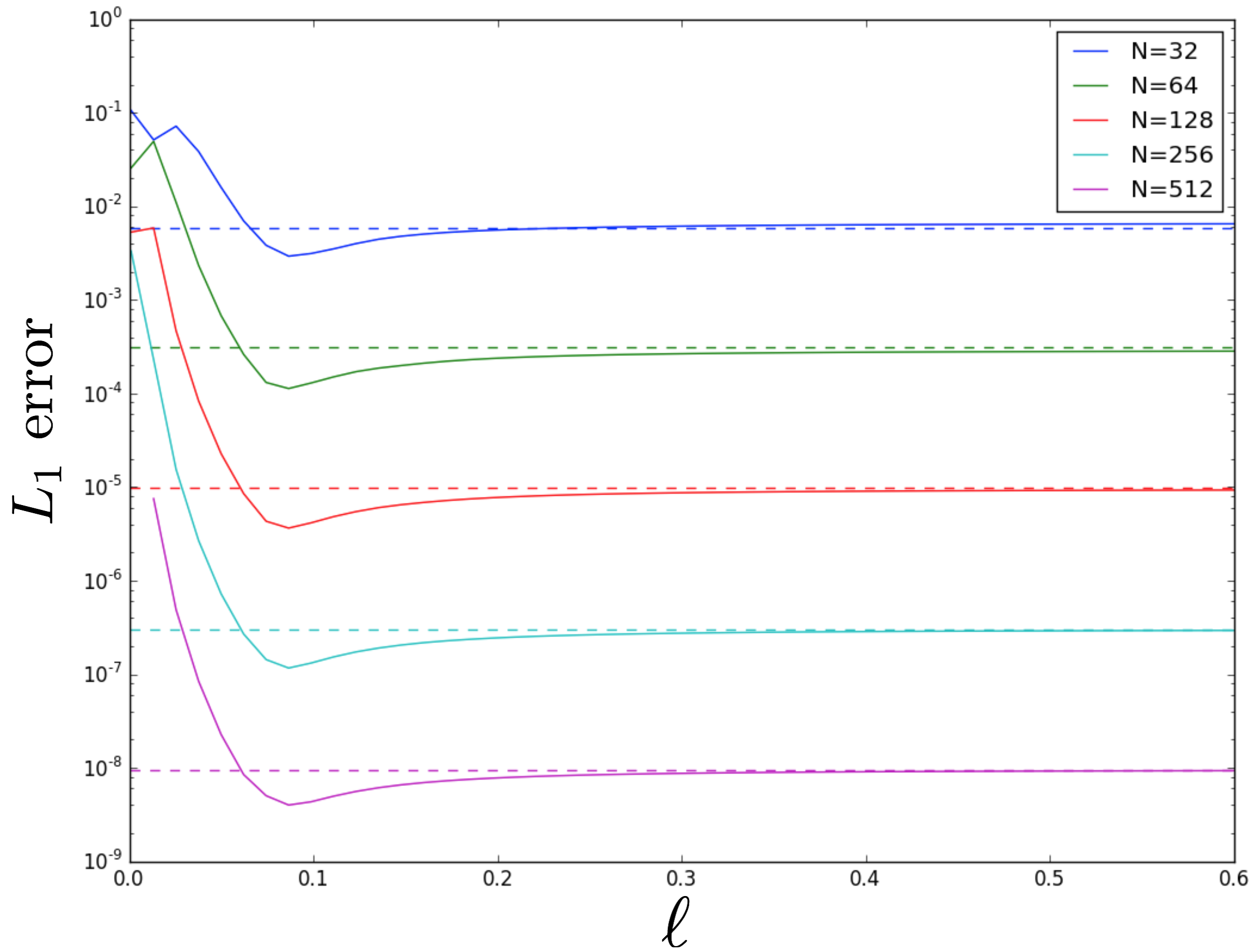}}
  \caption{(a) Convergence for smooth Gaussian density advection using
    different GP stencil radii, all using the length scale $\ell=0.1$
    and the HLLC Riemann solver with $C_{\text{cfl}}=0.8$.
  Black dotted lines  show $2R+1$ convergence rates, ranging from 3rd-order ($R=1$) to 11th-order ($R=5$). 
  Red dotted line represents a plateau at an $L_1$ of $10^{-12}$ where a large condition number of the covariance matrix is obtained 
  and no further accuracy is achievable.
  (b) $L_1$ errors as a function of the hyperparameter $\ell$, using $R=2$. Dotted lines are the error for
fifth-order WENO-JS method on the same stencil.}
  \label{fig:conv}
\end{figure}

The test considered here involves the passive advection of a Gaussian density profile. We initialize a computational box 
on [0,1] with periodic boundary conditions. 
The initial density profile is defined by $\rho(x) = 1 + e^{-100(x-x_0)^2}$, with $x_0=0.5$, with constant velocity, $u=1$, and 
pressure, $P=1/\gamma$. The specific heat ratio is chosen to be $\gamma=5/3$. 
The resulting profile is propagated for one period through the boundaries. At $t=1$, the profile returns to its 
initial position at $x=x_0$, any deformation of the initial profile is due to either phase errors or numerical diffusion. 
We perform this test using a length hyperparameter of $\ell=0.1$ for stencil radii $R=1,2,3,4$ and $5$, 
with a fixed Courant number, $C_{\text{cfl}}=0.8$ and vary the resolution of computational box, with $N=32,64,128,256$ and $512$.

The results of this study are shown in Fig.~\ref{conv:plot}. From these numerical experiments, GP reconstruction shows 
a convergence rate that goes as the size of the stencil, $2R+1$. 
Note that this convergence rate $2R+1$ of GP is equivalent to the convergence rate $2r-1$ of a classic WENO method
for a same size of stencil. 
In the classic WENO method, the notation $r$ represents the order (not degree) of polynomials on $S_m$
which require $r$ cells on each $S_m$, totaling $2r-1$ cells on $S_R$. This implies $2r-1=2R+1$, or $R=r-1$.
For example GP-R2 converges at fifth-order on the stencil $S_R=S_2$ that has five cells, 
so does WENO-JS with three third-order ENO polynomials (i.e., $r=3$) on the same 5-point stencil.

In Fig.~\ref{conv:plot} the $L_1$ error plateaus out at an  $\sim 10^{-12}$ which is
a few orders of magnitude greater than the standard IEEE double-precision, $\sim 10^{16}$.
This happens because at high resolution the length hyperparameter, $\ell$, becomes very large relative to the grid spacing, $\Delta$. 
The covariance matrix, $\mathbf{C}$ given in Eq. (\ref{eq:SE-cov}) 
becomes nearly singular in the regime $\ell/\Delta \gg 1$, yielding very large condition 
numbers for $\mathbf{C}$. We find the plateau in the $L_1$ error occurs for condition numbers, $\kappa \sim 10^{18}$, corresponding to the 
point where the errors in inverting $\mathbf{C}$ in Eq. (\ref{eq:vol_prediction}) begin to dominate. 
This implies that the choice of floating-point precision has an immense impact on the possible $\ell/\Delta$,
and a proper floating-point precision needs to be chosen in such a way that the condition number errors do not dominate.
As mentioned in Section \ref{sec:gp_se}, 
SE suffers from singularity when the size of the dataset grows.
The approach that enabled to produce the results in Fig.~\ref{fig:conv} 
is to utilize quadruple-precision
\textit{only} for the calculation of 
$\mathbf{z}^T=\mathbf{T}_*^{T}\mathbf{C}^{-1}$ in Eq. (\ref{eq:vol_prediction}).
Otherwise, the plateau would appear at a much higher $L_1$ error $\sim 10^{-7}$ with
double-precision, producing undesirable outcomes for all forms of grid
convergence studies.
This corresponds to condition numbers $\kappa\sim 10^8$, and as a
point of reference, the breakdown starts to occur for $\ell/\Delta>48$ using a GP radius $R=2$.
Since $\mathbf{z}^T$ needs to be calculated only once, before starting the simulation, 
it can then be truncated to double-precision for use in the actual reconstruction procedure. 
There are only four related small subroutines that need to be compiled with quadruple-precision
in our code implementation. The overall performance is not affected due to this extra
precision handling.
It should be noted that this extra precision handling 
is necessary for the purpose of a grid convergence study from the perspective of CFD applications.

The correlational length hyperparameter $\ell$ provides an additional avenue to tune
solution accuracy that is not present in polynomial-based
methods. Fig.~\ref{conv:length} shows how the GP errors with $R=2$ in the
smooth-advection problem changes with the choice of $\ell$, compared with
the error from a fifth-order WENO-JS + RK4 solution (denoted in dotted lines).
At large $\ell$ the errors
become roughly the same as in WENO-JS, and at small $\ell$ the errors become
worse. The error finds a minimum at a value of $\ell$ near the half-width
of the Gaussian density profile. 
This is in line with the idea that the optimal
choice of $\ell$ should match the physical length scale of the feature
being resolved. We will report new strategies on this topic in our future papers.
%

\subsection{1D Shu-Osher Shock Tube Problem}
\label{sec:shu-osher-shock}

\begin{figure}[htbp!]  
  \centering
    \subfigure[][]{
      \label{shu:comp}
      \includegraphics[width=0.8\textwidth]{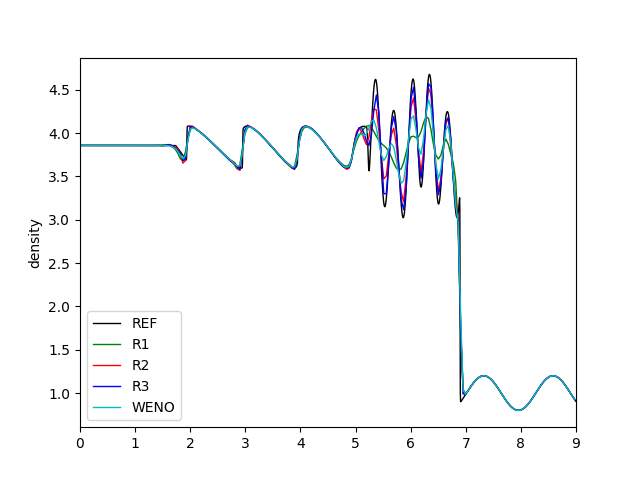}}
    \subfigure[][]{
      \label{shu:zoom}
      \includegraphics[width=0.8\textwidth]{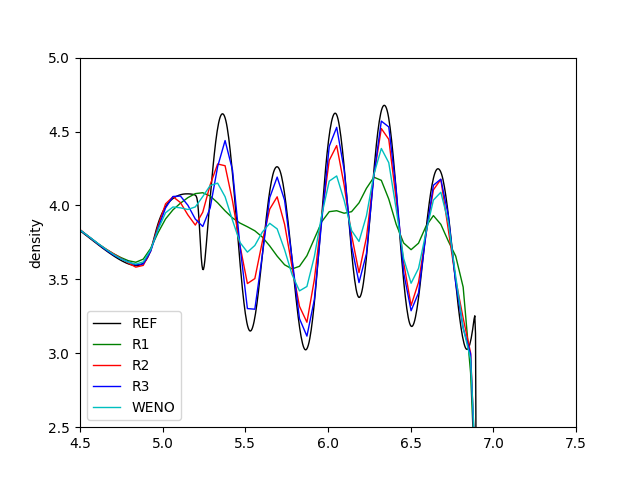}}
  \caption{The Shu-Osher problem. Density profiles at $t=1.8$ computed
    using three different GP stencil radii ($R=1,2,3$) on a 200 grid
    resolution with $C_{\text{cfl}}=0.8$. All GP calculations use
    $\ell/\Delta = 6$ and the HLLC Riemann solver. 
    The reference solution is obtained using WENO-JS on a 2056 grid resolution. 
    \ref{shu:zoom} shows a closeup view of the post shock features. }
  \label{fig:shu}  
\end{figure}

\begin{figure}[htbp!]  
  \centering
    \subfigure[][]{
      \label{shu:l3}
      \includegraphics[width=0.5\textwidth]{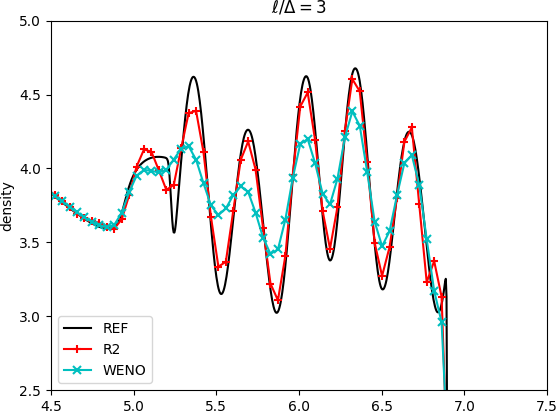}}
     \subfigure[][]{
      \label{shu:l6}
      \includegraphics[width=0.5\textwidth]{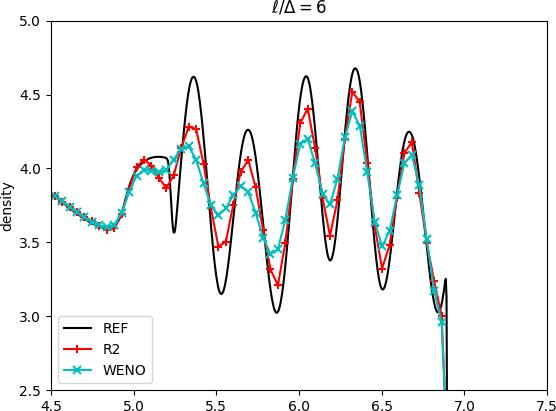}}
    \subfigure[][]{
      \label{shu:l24}
      \includegraphics[width=0.5\textwidth]{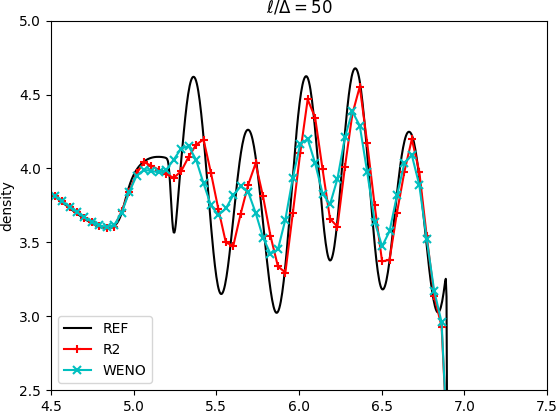}}
  \caption{Closeup view of the post shock features as in
    Fig.~\ref{shu:zoom} using $N=200$ for WENO-JS and GP-R2 (a) with
    $\ell/\Delta=3$, (b) with $\ell/\Delta=6$, and (c) with $\ell/\Delta=50$. The reference
    solution is calculated using WENO-JS on 2056 grid resolution. }
  \label{fig:shu-ldel}  
\end{figure}

The second test is the Shu-Osher problem~\cite{Shu1989} to test 
GP's shock-capturing capability as well as to see how well GP can resolve small-scale features in the flow.
The test gives a good indication 
of the method's numerical diffusivity, and it has been a popular benchmark to 
demonstrate numerical errors of a given method. In this problem, 
a (nominally) Mach 3 shock wave propagates into a constant density field with 
sinusoidal perturbations. As the shock advances, 
two sets of density features appear behind the shock. 
One set has the same spatial frequency 
as the unshocked perturbations, while in the second set 
the frequency is doubled and follows more closely behind the shock. 
The primary test of the numerical method is to accurately 
resolve the dynamics and strengths of the oscillations behind the shock.

The results are shown in Fig.~\ref{fig:shu}. The solutions are calculated at $t=1.8$ using a 
resolution of $N=200$ and are compared to a reference solution
resolved on $N=2056$. It is evident that the GP solution using $R=3$
provides the least diffusive solution of the methods shown, especially
in capturing the amplitude of the post-shock oscillations in
Fig.~\ref{shu:zoom}. Of the two fifth-order methods, GP-R2 and WENO-JS,
the GP solution has slightly better amplitude in the post-shock
oscillations compared to the WENO-JS solution, consistent with what is observed in
Section~\ref{sec:smooth-advection} for the smooth advection
problem. 

The results in Section~\ref{sec:smooth-advection} suggest
that the choice of $\ell$ should correspond with a length
scale characteristic of the flow for optimal
performance. Fig.~\ref{fig:shu-ldel} compares the post shock features
on the same grid for the WENO-JS method and the GP-R2 method
with $\ell/\Delta=3, 6$ and $50$. Here $\ell/\Delta=3$
is roughly a half wavelength of the oscillations, and the GP-R2
method clearly gives a much more accurate solution compared to the WENO-JS
solution.  Just as can be seen in Fig.~\ref{conv:length}, $\ell/\Delta$
much larger than the characteristic length yields a solution much
closer to that of WENO-JS.
From Fig.~\ref{conv:length} we expect that the GP reconstructions in
Eq.~(\ref{eq:Sm_prediction}) should approach the WENO-JS
reconstructions in smooth regions for $\ell \gg \Delta$. However, we
see that the new GP based smoothness indicators allow for the
amplitudes near the shock to be better resolved. 
%
This reflects a key advantage of the proposed GP method
over polynomial-based high-order methods. The additional flexibility afforded by the
kernel hyperparameters in GP, allowing for solution accuracy to be
tuned to the features that are being resolved. Only at larger values
of $\ell$ does the model
become fully constrained by the data in GP, whereas the interpolating
polynomials used in a classical WENO are always fully constrained by design. 
Analogously in the RBF theory, the 
the shape parameters $\epsilon_j$
plays an important role in the similar context of accuracy and convergence.
Several strategies have been studied recently for solving hyperbolic PDEs
\cite{bigoni2016adaptive,guo2017rbf,jung2010recovery}.

\subsection{The Sod Shock Tube Test}
\label{sec:sod-shock-tube}

\begin{figure}[h]
\centering
\begin{tabular}{cc}
\subfigure[][]{
\includegraphics[width=0.5\textwidth]{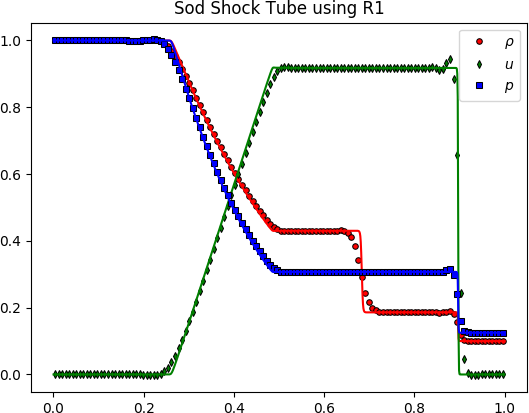}}
\subfigure[][]{
\includegraphics[width=0.5\textwidth]{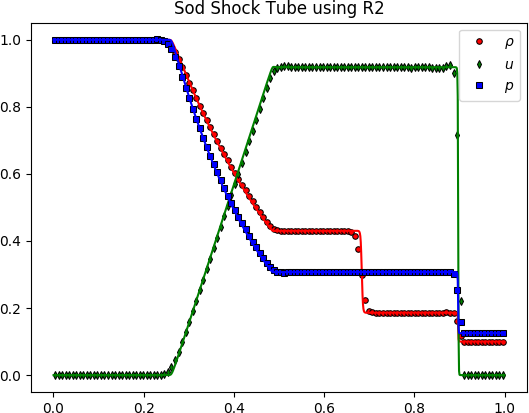}}\\
\subfigure[][]{
\includegraphics[width=0.5\textwidth]{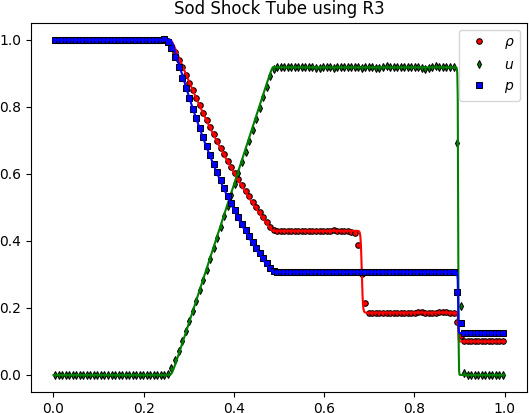}}
\subfigure[][]{
\includegraphics[width=0.5\textwidth]{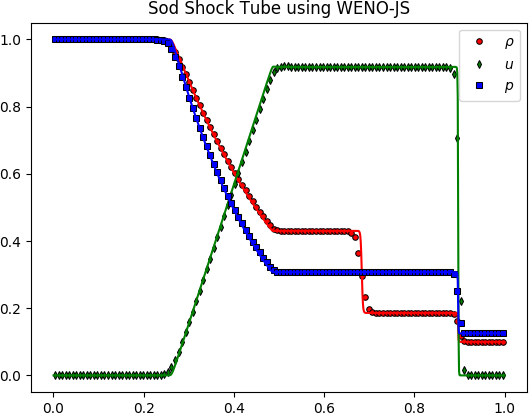}}\\
\end{tabular}
\caption{The Sod shock tube problem at $t=0.2$.
(a) GP with stencil radius $R=1$, (b) GP with stencil radius $R=2$, (c) GP with stencil radius $R=3$,
and (d) WENO-JS,  all using 128 grid points with $C_{\text{cfl}}=0.8$. All GP calculations use
$\ell/\Delta = 12$ and the HLLC Riemann solver. 
Solid lines show a reference solution computed using WENO-JS on 1024 grid points.}
\label{Fig:Sod}
\end{figure}

The shock tube problem of Sod \cite{sod1978survey} has been one of the
most popular 1D tests of a code's ability to handle shocks and contact
discontinuities. The initial conditions, on the domain $[0,1]$, are
given by the left and right states
\begin{equation}
  \label{eq:sod}
  (\rho, u, p) = \left \{
    \begin{array}{lr}
      (1,0,1), & x<0.5, \\
      (0.125,0,0.1), & x>0.5,
    \end{array}
    \right .
\end{equation}
with the ratio of specific heats $\gamma=1.4$. Outflow boundary
conditions are imposed at $x=0$ and $x=1$. The results 
are shown in Fig.~\ref{Fig:Sod}. 

All of the GP-WENO schemes correctly
predicts the nonlinear characteristics of the flow including the rarefaction
wave, contact discontinuity and the shock. The solution using GP-R2
is very comparable to the WENO-JS solution using the same 5-point stencil.
As expected, the GP-R1 solution smears out the most at both the shock
and the contact discontinuity, and at the head and tail of the rarefaction.
The 7th order GP-R3 also successfully demonstrates that its shock solution
is physically correct without triggering any unphysical
oscillation. Somewhat counter intuitively from the perspective of 1D
polynomial schemes, the smallest stencil GP-R1 shows the most
oscillations near the shock. This happens because the
eigendecomposition of the kernel in Eq.~(\ref{eq:matrix_eigen}) used in
the calculation the smoothness indicators for Eq.~(\ref{eq:betas})
better approximates $\|f\|_\mathcal{H}^2$ as the size of the stencil
increases. Recall that the eigenexpansion of the kernel function in
Eq.~(\ref{eq:Efcn}) can contain an infinite number of eigenvalues, and
so can $\|f\|_\mathcal{H}^2$ and the sum in
Eq.~(\ref{eq:Hnorm}). The finite approximation to the eigensystem and
subsequently $\|f\|_\mathcal{H}^2$ then becomes better as the smallest
coefficient $\alpha_i$ goes to zero. Then for GP, the $\beta_m$'s best
indicate the smoothness on larger sized stencils. 


\subsection{The Einfeldt Strong Rarefaction Test}
\label{sec:einf-strong-raref}

\begin{figure}[h!]
\centering
\begin{tabular}{cc}
\subfigure[][]{
\includegraphics[width=0.49\textwidth]{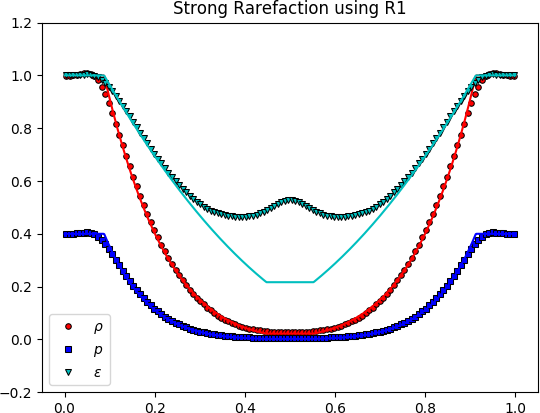}}
& \subfigure[][]{
\includegraphics[width=0.49\textwidth]{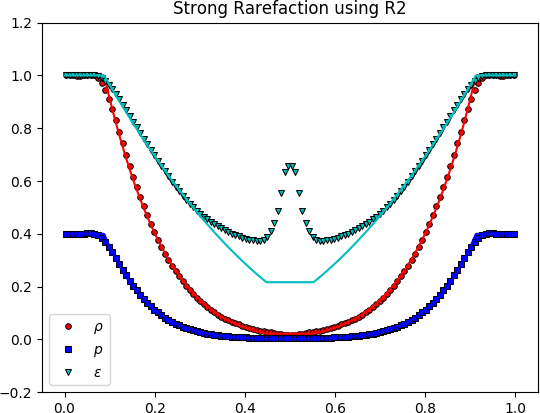}}\\
\subfigure[][]{
\includegraphics[width=0.49\textwidth]{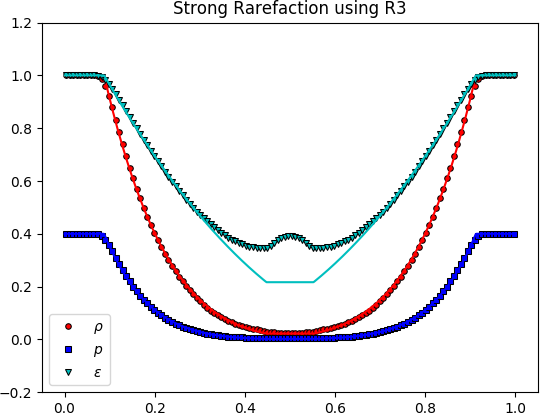}}
& \subfigure[][]{
\includegraphics[width=0.49\textwidth]{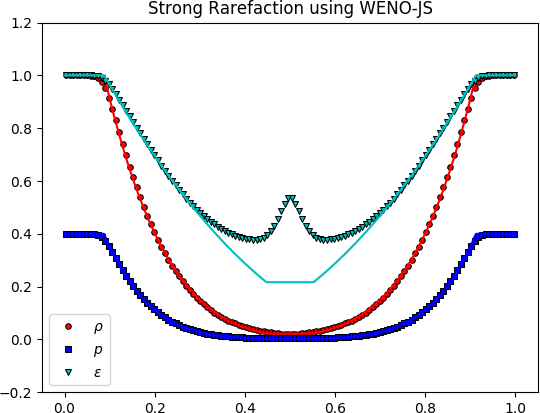}}\\
\end{tabular} 
\caption{The Einfeldt strong rarefaction test at $t=0.15$.
(a) GP with stencil radius $R=1$, (b) GP with stencil radius $R=2$, (c) GP with stencil radius $R=3$,
and (d) WENO-JS, all using 128 grid points with
$C_{\text{cfl}}=0.8$. All GP calculations use $\ell/\Delta = 12$ and
the HLLC Riemann solver. Solid lines show the exact solution.}
\label{fig:einfeldt}
\end{figure}

First described by Einfeldt et al. \cite{einfeldt1991godunov}, this
problem tests how satisfactorily a code can perform in a low density
region in computing physical variables, $\rho,u,p,\epsilon$,
etc. Among those, the internal energy $\epsilon=p/(\rho(\gamma-1))$ is
particularly difficult to get right due to regions where the density and pressure
are very close to zero. The ratio of these two small values 
amplifies any small errors in both $\rho$ and $p$, making the errors
in $\epsilon$ the largest in general \cite{toro2009}. The large errors
in $\epsilon$ are apparent for all schemes shown in
Fig.~\ref{fig:einfeldt}, where the error is largest around $x=0.5$. 
It can be observed that the amount of departure in $\epsilon$ 
from the exact solution (the cyan solid line) decreases 
as the GP radius $R$ increases. The error in GP-R2 is slightly larger
for $\epsilon$ at the center than in WENO-JS. However, the peak becomes
considerably smaller in amplitude and becomes slightly
flatter as $R$ increases. 

\subsection{Brio-Wu MHD Shock Tube}
\label{sec:brio-wu-mhd}

\begin{figure}[h!]
\centering
\begin{tabular}{cc}
\subfigure[][]{
\includegraphics[width=0.49\textwidth]{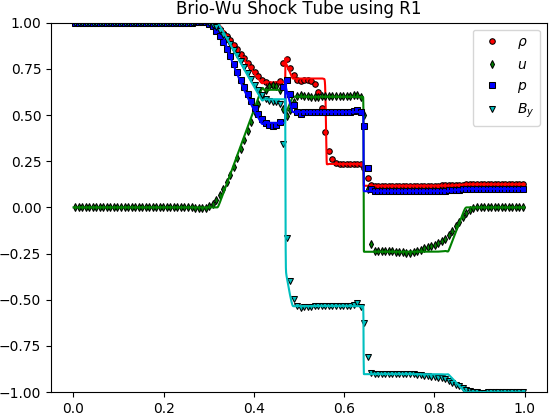}}
\subfigure[][]{
\includegraphics[width=0.49\textwidth]{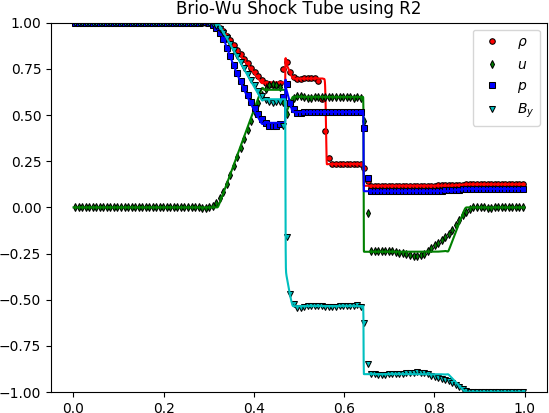}}\\
\subfigure[][]{
\includegraphics[width=0.49\textwidth]{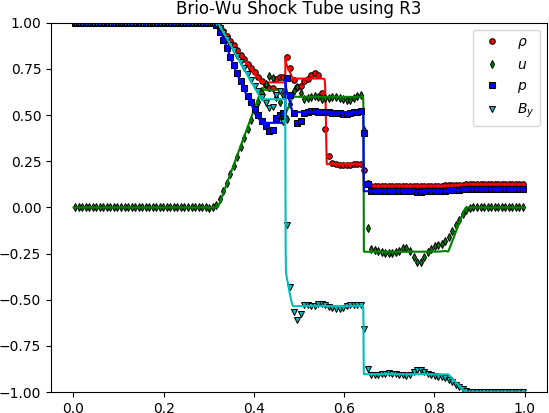}}
\subfigure[][]{
\includegraphics[width=0.49\textwidth]{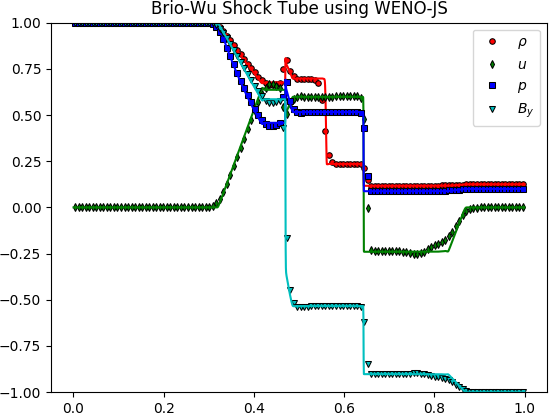}}\\
\end{tabular} 
\caption{The Brio-Wu shock tube at $t=0.1$.
(a) GP with stencil radius $R=1$, (b) GP with stencil radius $R=2$, (c) GP with stencil radius $R=3$,
and (d) WENO-JS, all using 128 grid points with
$C_{\text{cfl}}=0.8$. All GP calculations use $\ell/\Delta = 12$ and
the Roe Riemann solver. Solid lines show a reference solution computed using WENO-JS on 1024 grid points.}
\label{fig:brio-wu}
\end{figure}

Brio and Wu \cite{brio1988upwind} studied an MHD version of Sod's
shock tube problem, which has since become an essential test for any
MHD code. The test has since uncovered some interesting findings, such
as the compound wave \cite{brio1988upwind}, as well as the existence of
non-unique solutions \cite{torrilhon2003non,torrilhon2003uniqueness}. 
The results for this test are shown in
Fig.~\ref{fig:brio-wu}. 

All of the GP methods are able to
satisfactorily capture the MHD structures of the problem. In all methods,
including WENO-JS, there are some observable oscillations in the post
shock regions. Lee in \cite{lee2011upwind} showed that these
oscillations arise as a result of the numerical nature of the slowly
moving shock \cite{woodward1984numerical} 
controlled by the strength of the transverse magnetic field. 
As studied by various researchers 
\cite{arora1997postshock,jin1996effects,johnsen2008numerical,karni1997computations,roberts1990behavior,stiriba2003numerical},
there seems no ultimate fix for controlling such unphysical oscillations due to
the slowly moving shock. Quantitatively, the amount of oscillations differs
in different choices of numerical methods such as reconstruction algorithms and Riemann solvers.
We see that all of the GP solutions together with the WENO-JS solution feature a comparable
level of oscillations. Except for GP-R1, all solutions also suffer from a similar type
of distortions in $u$ and $B_y$ in the right going fast rarefaction. This distortion
as well as the oscillations due to the slowly moving shock seem to be suppressed
in the 3rd order GP-R1.

\subsection{Ryu and Jones MHD Shock Tubes}
\label{sec:ryu-jones-mhd}

Ryu and Jones \cite{ryu1994numerical} introduced a large set of MHD
shock tube problems as a test of their 1D algorithm, that are now
informative to run as a code verification. In what follows we will
refer to the tests as RJ followed by the corresponding figure from
\cite{ryu1994numerical} in which the test can be found.

\subsubsection{RJ1b Shock Tube}
\label{sec:rj1b-shock-tube}

\begin{figure}[h!]
\centering
\begin{tabular}{cc}
\subfigure[][]{
\includegraphics[width=0.49\textwidth]{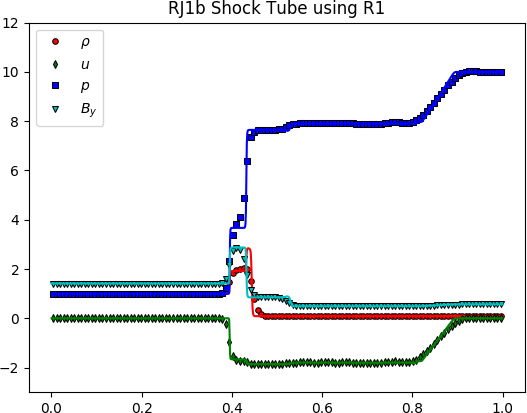}}
\subfigure[][]{
\includegraphics[width=0.49\textwidth]{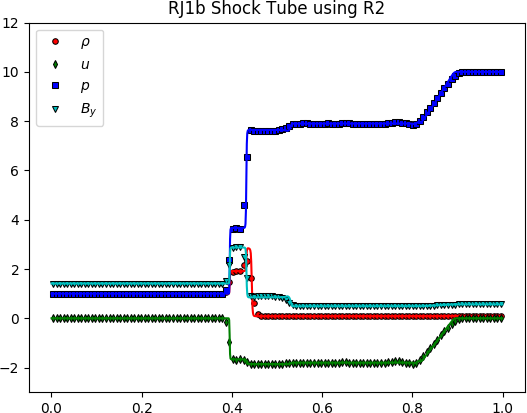}}\\
\subfigure[][]{
\includegraphics[width=0.49\textwidth]{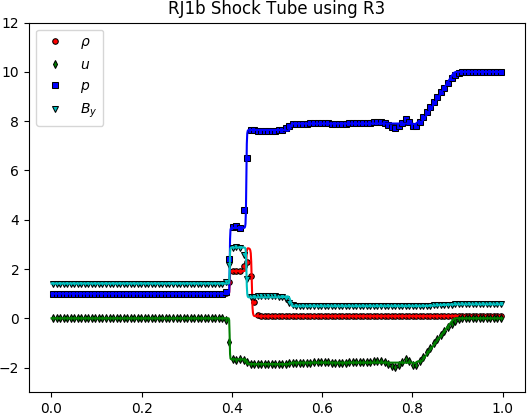}}
\subfigure[][]{
\includegraphics[width=0.49\textwidth]{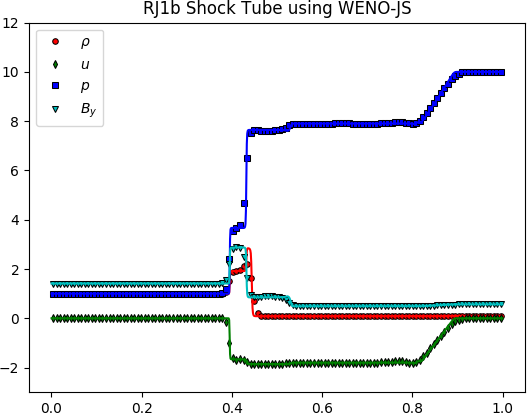}}\\
\end{tabular} 
\caption{The RJ1b MHD shock tube at $t=0.03$.
(a) GP with stencil radius $R=1$, (b) GP with stencil radius $R=2$, (c) GP with stencil radius $R=3$,
and (d) WENO-JS, all using 128 grid points with $C_{\text{cfl}}=0.8$. 
All GP calculations use $\ell/\Delta = 12$ and the Roe Riemann solver. 
Solid lines show a reference solution computed using WENO-JS on 1024 grid points. }
\label{fig:rj1b}
\end{figure}

The first of the RJ shock tubes we consider is the RJ1b problem. This
test contains a left going fast and slow shock, contact discontinuity
as well as a slow and fast rarefaction. In Fig.~\ref{fig:rj1b} that
all waves are resolved in the schemes considered.

\subsubsection{RJ2a Shock Tube}
\label{sec:rj2a-shock-tube}

\begin{figure}[h!]
\centering
\begin{tabular}{cc}
\subfigure[][]{
\includegraphics[width=0.49\textwidth]{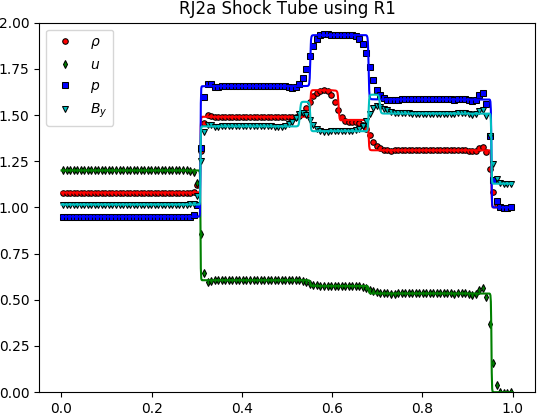}}
\subfigure[][]{
\includegraphics[width=0.49\textwidth]{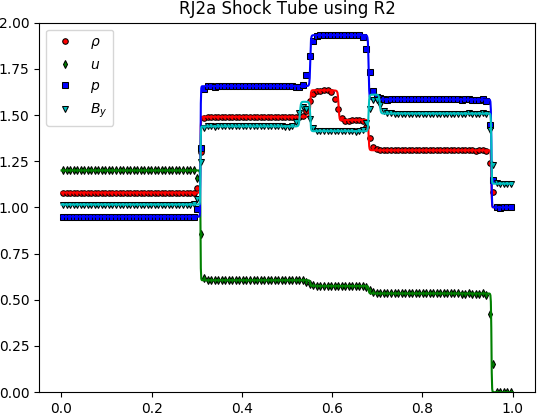}}\\
\subfigure[][]{
\includegraphics[width=0.49\textwidth]{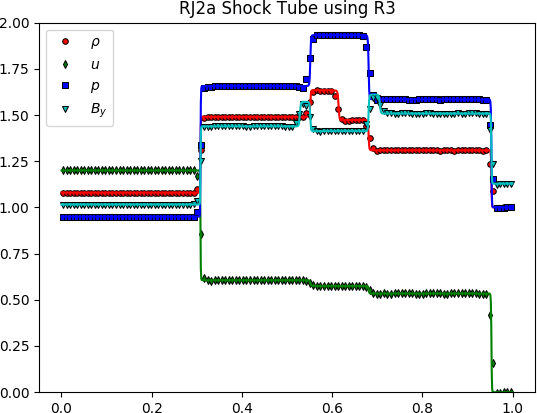}}
\subfigure[][]{
\includegraphics[width=0.49\textwidth]{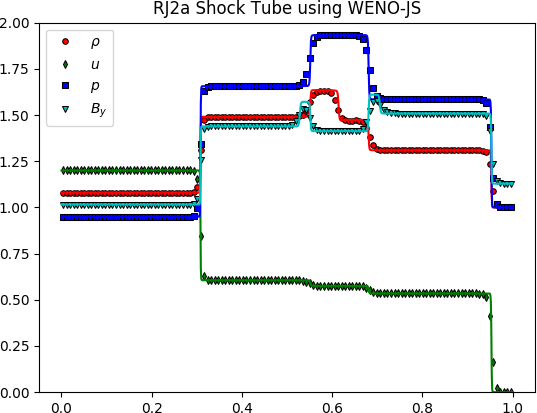}}\\
\end{tabular} 
\caption{The RJ2a MHD shock tube at $t=0.03$.
(a) GP with stencil radius $R=1$, (b) GP with stencil radius $R=2$, (c) GP with stencil radius $R=3$,
and (d) WENO-JS, all using 128 grid points with $C_{\text{cfl}}=0.8$. 
All GP calculations use $\ell/\Delta = 12$ and the HLLC Riemann solver. 
Solid lines show a reference solution computed using WENO-JS on 1024 grid points. }
\label{fig:rj2a}
\end{figure}

The RJ2a test provides an interesting test due to its initial
conditions producing a discontinuity in each of the MHD wave
families. The solution contains both fast- and slow- left and
right-moving magnetoacoustic shocks, left- and right-moving rotational
discontinuities and a contact discontinuity. Fig.~\ref{fig:rj2a} shows
that all three of the GP schemes are able to resolve all of these
discontinuities. Again we see that the smallest stencil GP-R1 solution
contains some oscillations near the shock that are not present in the
other GP solutions.

\subsubsection{RJ2b Shock Tube}
\label{sec:rj2b-shock-tube}

\begin{figure}[h!]
\centering
\begin{tabular}{cc}
\subfigure[][]{
\includegraphics[width=0.49\textwidth]{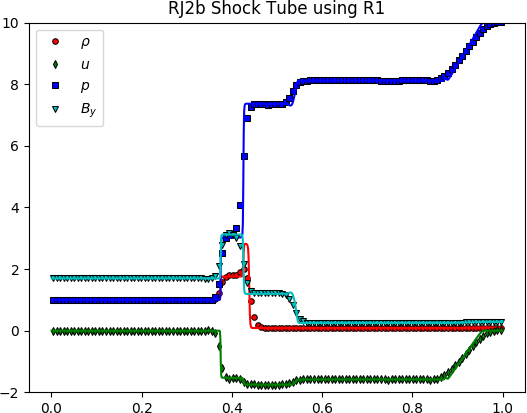}}
\subfigure[][]{
\includegraphics[width=0.49\textwidth]{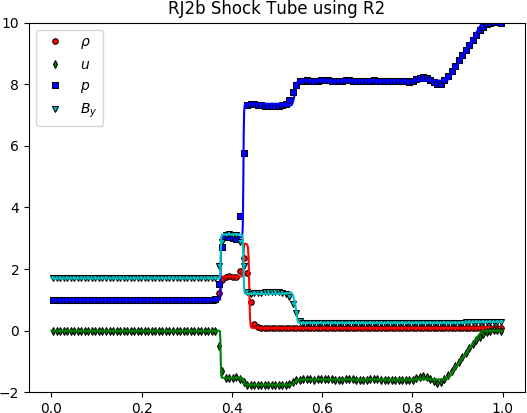}}\\
\subfigure[][]{
\includegraphics[width=0.49\textwidth]{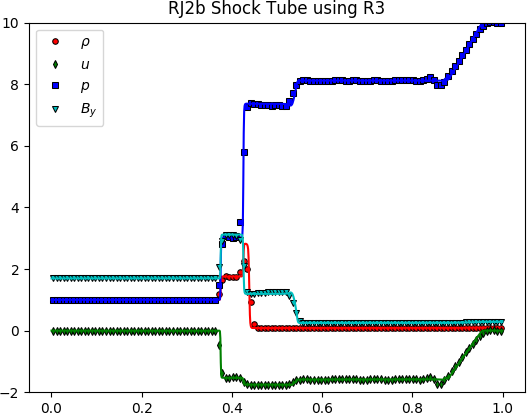}}
\subfigure[][]{
\includegraphics[width=0.49\textwidth]{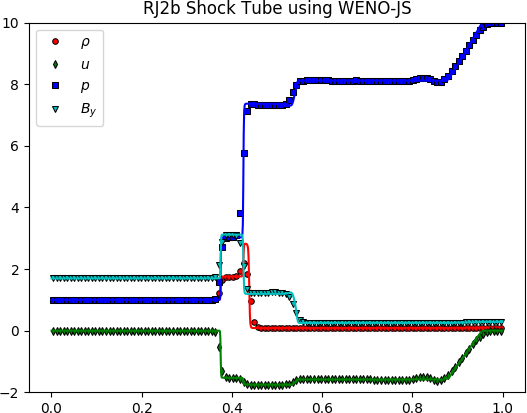}}\\
\end{tabular} 
\caption{The RJ2b MHD shock tube at $t=0.2$.
(a) GP with stencil radius $R=1$, (b) GP with stencil radius $R=2$, (c) GP with stencil radius $R=3$,
and (d) WENO-JS, all using 128 grid points with $C_{\text{cfl}}=0.8$. 
All GP calculations use $\ell/\Delta = 12$ and the Roe Riemann solver. 
Solid lines show a reference solution computed using WENO-JS on 1024 grid points. }
\label{fig:rj2b}
\end{figure}

The RJ2b shock tube creates a set of a fast shock, a rotational
discontinuity and a slow shock moving to the left away from a contact
discontinuity, as well as a fast rarefaction, rotational discontinuity
and slow rarefaction all moving to the right. What is of interest is
that since the waves propagate at almost the same speed, at $t=0.035$
they have still yet to separate much, and so test a codes ability to
resolve all of the discontinuities despite their close separation. The
results of this test for the methods considered are shown in
Fig.~\ref{fig:rj2b}. At the shown resolution of $N=128$, the contact
discontinuity and the slow shock become somewhat smeared together in
the GP-R1 solution, while they are better resolved for the other
methods, even though there are only couple of grid points distributed
over the range of the features.

\subsubsection{RJ4a Shock Tube}
\label{sec:rj4a-shock-tube}

\begin{figure}[h!]
\centering
\begin{tabular}{cc}
\subfigure[][]{
\includegraphics[width=0.49\textwidth]{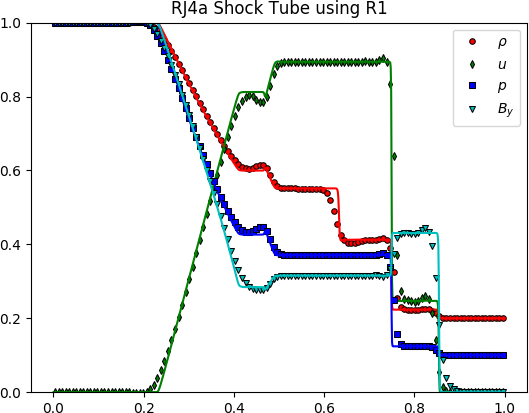}}
\subfigure[][]{
\includegraphics[width=0.49\textwidth]{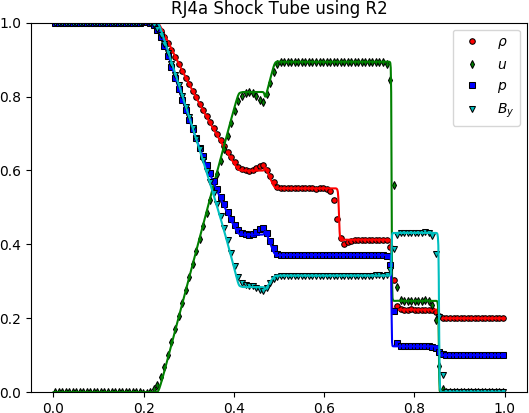}}\\
\subfigure[][]{
\includegraphics[width=0.49\textwidth]{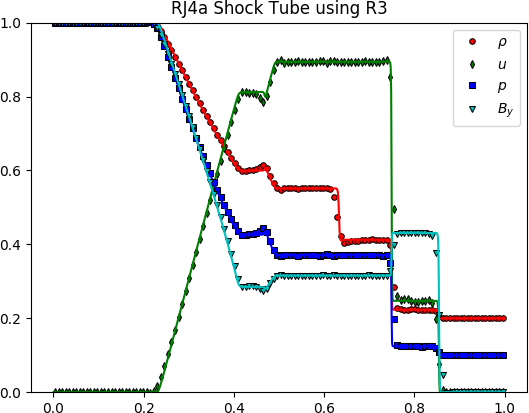}}
\subfigure[][]{
\includegraphics[width=0.49\textwidth]{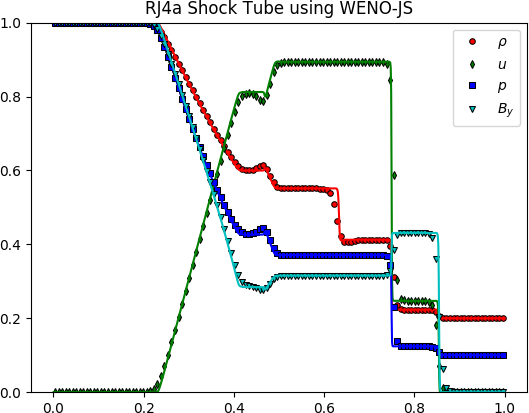}}\\
\end{tabular} 
\caption{The RJ4a MHD shock tube at $t=0.15$.
(a) GP with stencil radius $R=1$, (b) GP with stencil radius $R=2$, (c) GP with stencil radius $R=3$,
and (d) WENO-JS, all using 128 grid points with $C_{\text{cfl}}=0.8$. 
All GP calculations use $\ell/\Delta = 12$ and the HLLC Riemann solver. 
Solid lines show a reference solution computed using WENO-JS on 1024 grid points. }
\label{fig:rj4a}
\end{figure}

The RJ4a shock tube yields a fast and slow rarefaction, a contact
discontinuity, a slow shock, and of particular interest the switch-on
fast shock. The feature of the switch-on shock is that the magnetic
field turns on in the region behind the shock. As can be seen in
Fig.~\ref{fig:rj4a}, all of the features including the switch-on fast
shock are resolved in all methods. 
We see that GP-R1 smears out the solution not only 
in resolving discontinuous flow regions,
but also in resolving both fast and slow rarefaction waves.

\subsubsection{RJ4b Shock Tube}
\label{sec:rj4b-shock-tube}

\begin{figure}[h!]
\centering
\begin{tabular}{cc}
\subfigure[][]{
\includegraphics[width=0.49\textwidth]{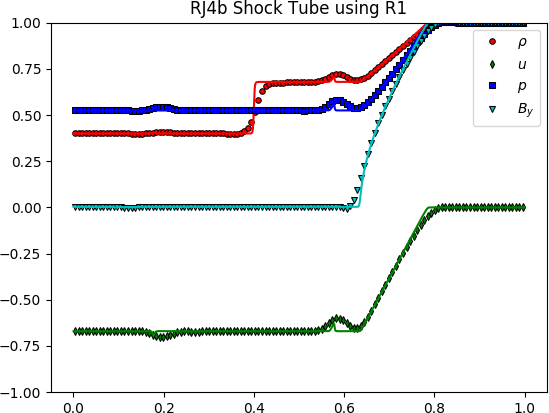}}
\subfigure[][]{
\includegraphics[width=0.49\textwidth]{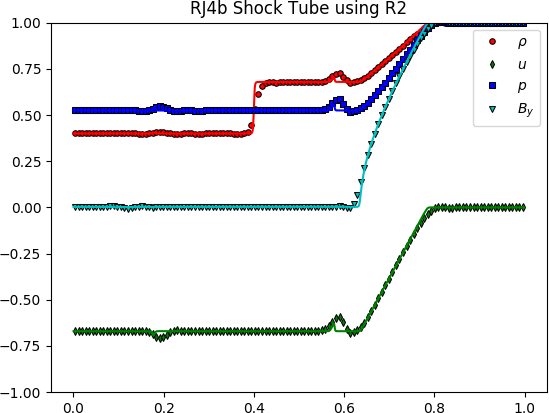}}\\
\subfigure[][]{
\includegraphics[width=0.49\textwidth]{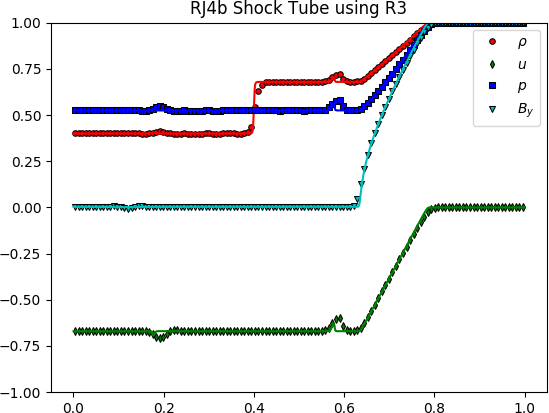}}
\subfigure[][]{
\includegraphics[width=0.49\textwidth]{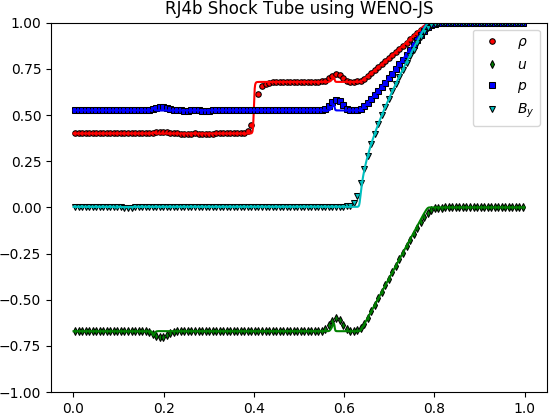}}\\
\end{tabular} 
\caption{The RJ4b MHD shock tube at $t=0.15$.
(a) GP with stencil radius $R=1$, (b) GP with stencil radius $R=2$, (c) GP with stencil radius $R=3$,
and (d) WENO-JS, all using 128 grid points with $C_{\text{cfl}}=0.8$. 
All GP calculations use $\ell/\Delta = 12$ and the HLLC Riemann solver. 
Solid lines show a reference solution computed using WENO-JS on 1024 grid points. }
\label{fig:rj4b}
\end{figure}

The RJ4b test is designed to produce only a contact discontinuity and a fast right going
switch-off fast rarefaction, where the magnetic field is zero behind
the rarefaction. We can see in Fig.~\ref{fig:rj4b} that both the
contact discontinuity and the switch-off rarefaction features are
captured in all of the considered methods.

\subsubsection{RJ5b Shock Tube}
\label{sec:rj5b-shock-tube}

\begin{figure}[h!]
\centering
\begin{tabular}{cc}
\subfigure[][]{
\includegraphics[width=0.49\textwidth]{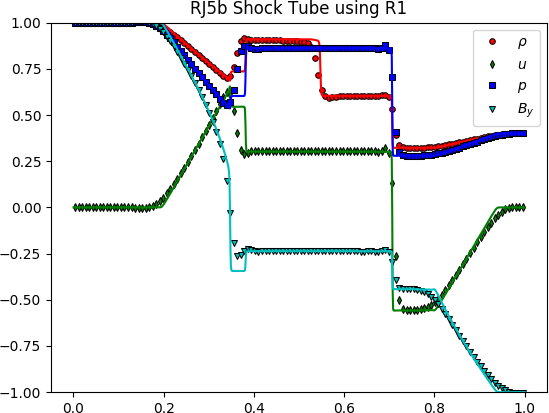}}
\subfigure[][]{
\includegraphics[width=0.49\textwidth]{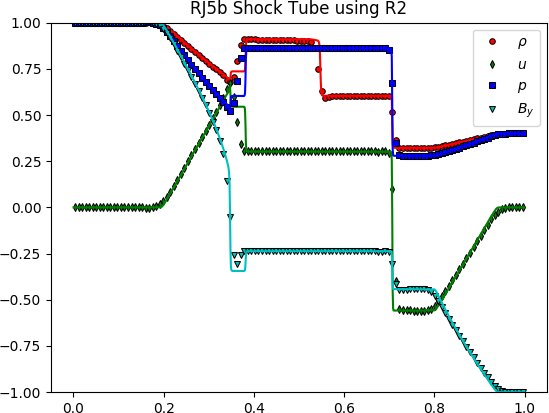}}\\
\subfigure[][]{
\includegraphics[width=0.49\textwidth]{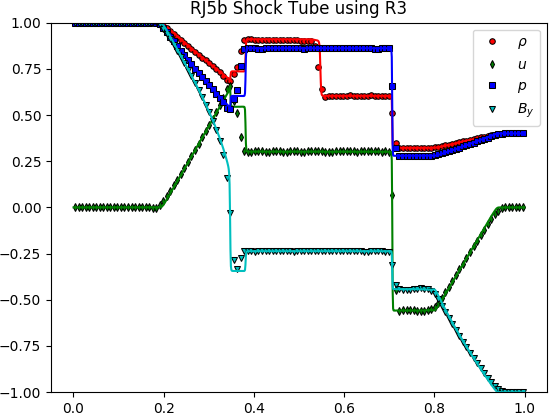}}
\subfigure[][]{
\includegraphics[width=0.49\textwidth]{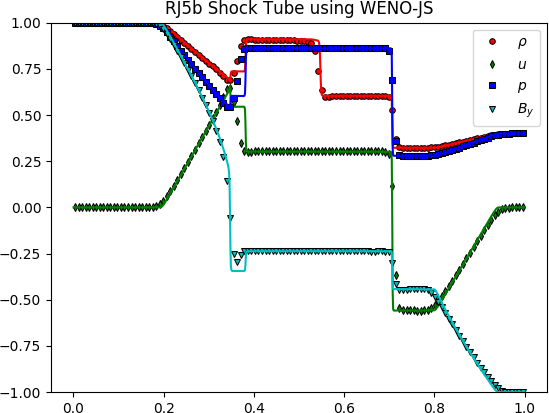}}\\
\end{tabular} 
\caption{The RJ5b MHD shock tube at $t=0.16$.
(a) GP with stencil radius $R=1$, (b) GP with stencil radius $R=2$, (c) GP with stencil radius $R=3$,
and (d) WENO-JS, all using 128 grid points with $C_{\text{cfl}}=0.8$. 
All GP calculations use $\ell/\Delta = 12$ and the HLLC Riemann solver. 
Solid lines show a reference solution computed using WENO-JS on 1024 grid points.}
\label{fig:rj5b}
\end{figure}

The RJ5b problem is of interest because it produces a fast compound
wave, as opposed to the slow compound wave in the Brio-Wu problem, in
addition to a left- and right-going slow shock, contact discontinuity
and fast rarefaction. At the resolution of $N=128$ used for the
results in Fig.~\ref{fig:rj5b}, the compound wave and one of the slow
shocks are smeared together in all the methods tested.

\section{Conclusion}
We summarize key novel features of the new high-order GP approach presented in this paper.

The new GP approach utilizes the key idea from statistical theory of GP prediction 
to produce accurate interpolations of fluid variables in CFD applications. 
We have developed a new set of numerical strategies of GP for both smooth flows and non-smooth
flows to numerically solve hyperbolic systems of conservation laws.

The GP methods presented here show an extremely fast rate of solution accuracy in 
smooth advection problems by controlling a single parameter, 
$R$. Further, the additional flexibility offered by the GP model
approach over the fully constrained polynomial based model
through the kernel hyperparameter $\ell$ allows for added tuning of
solution accuracy that is not present in traditional polynomial based high-order
methods. 
These parameters allow
the GP method to demonstrate variable orders of method accuracy as functions of
the size of the GP stencil and the hyperparameter $\ell$ (see Eq. (\ref{eq:SE}))
within a single algorithmic implementation.

The new GP based smoothness indicators introduced here are used to
construct non-linear weights that give the essentially non-oscillatory
property in discontinuous flows. The new smoothness indicators show a
significant advantage over traditional WENO schemes in capturing flow
features at the grid resolution near discontinuities.

The GP model, by design, can easily be extended to
multidimensional GP stencils. Therefore, GP can seamlessly provide
a significant algorithmic advantage in solving the multidimensional PDE of CFD.
This ``dimensional agnosticism'' is unique to GP, and not a feature of polynomial methods.
We will report our ongoing developments of GP in multiple spatial dimensions in forthcoming papers.


\section{Acknowledgements}
%
%
The software used in this work was developed in part
by funding from the U.S. DOE NNSA-ASC and OS-OASCR
to the Flash Center for Computational Science at the University
of Chicago.

\appendix

\section{Appendix: Choosing the Optimal Mean $f_0$ using the Maximum Likelihood Function}
\label{sec:appendix_meanFunction}
In most practical applications we will not have any prior information on the mean of the function samples.
This makes it reasonable to take a zero mean function $f_0=0$
if we try to achieve the simplest and most general symmetry of any random samples.
However, one can easily determine the best optimal choice of the mean value by maximizing the 
likelihood of the samples. 
This can be done by taking the logarithm function of Eq.~(\ref{eq:Like_1}) to first get the $\log$ likelihood function 
of the function samples $\mathbf{f}=[f(\bx_1), \dots, f(\bx_N)]^T$ that are pointwise,
\beq
\ln{\mathcal{L}}=\ln {P(\mathbf{f})} = 
-\frac{1}{2}\left(\mathbf{f}-\bar{\mathbf{f}}\right)^{T}\mathbf{K}^{-1}\left(\mathbf{f}-\bar{\mathbf{f}}\right)
-\frac{1}{2}\ln({\det|\mathbf{K}|})
-\frac{N}{2}\ln({2\pi}).
\label{eq:logLikelihood}
\eeq
To maximize the likelihood, we take the partial derivative of Eq.~(\ref{eq:logLikelihood}) with respect to $f_0$, 
where $\bff=f_0\mathbf{1}_N$,
and set the partial derivative to be zero,
\bea
0=\frac{\partial \mathcal{L}}{\partial f_0} 
&=& -\bff^T \bK\inv\mathbf{1}_N - \mathbf{1}_N^T \bK\inv\bff + 2 f_0\left( \mathbf{1}_N^T \bK\inv \mathbf{1}_N \right)\nonumber \\
&=& -2 \bff^T \bK\inv \mathbf{1}_N + 2 f_0\left( \mathbf{1}_N^T \bK\inv \mathbf{1}_N \right),
\eea
where we used the symmetry of $\bK\inv$ to get $\bff^T \bK\inv\mathbf{1}_N = \mathbf{1}_N^T \bK\inv\bff$.
The best optimal value of $f_0$ is then given by
\beq
f_0 = \frac{ \mathbf{1}_N^T \bK\inv\bff}{\mathbf{1}_N^T \bK\inv \mathbf{1}_N},
\label{eq:f_0_interpolation}
\eeq
which maximizes the likelihood of the samples in Eq.~(\ref{eq:Like_1}) for the case of interpolations.

In the similar way, we obtain the best optimal value of $f_0$ for the case of reconstructions, given as
\beq
f_0 = \frac{ \mathbf{1}_N^T \bC\inv\bG}{\mathbf{1}_N^T \bC\inv \mathbf{1}_N},
\label{eq:f_0_reconstruction}
\eeq
where $\bG=[\langle q_1\rangle, \dots, \langle q_N\rangle]^T$ contains $N$ volume-averaged values and $\bC$ is the integrated
kernel in Eq.~(\ref{eq:G_covariance}).

These optimal choices of $f_0$ are obviously more expensive in computation as they involve
local calculations on each GP stencil. In all the tests we presented 
in this paper we do not see any significant evidence that these optimal values result in any gains in terms of
solution accuracy and computational efficiency.


\section{Appendix: Operation Counts}
\label{sec:appendix_operationCounts}
We display possible outcomes of computing 
the GP-based new smoothness indicators 
in two different approaches described in Section \ref{sec:gp-weno}:

\bit
\item {\bf Option 1}: Compute $\alpha_i^m$ using Eqs. (\ref{eq:rec_fm}) and (\ref{eq:alphas_1}),
\item {\bf Option 2}: Compute $\alpha_i^m$ using Eqs. (\ref{eq:Pvec}) and (\ref{eq:alphas_2}).
\eit

Consider a 1D domain discretized with $N_x$ grid cells. Let's say we use a GP stencil of a radius $R$,
which is subdivided into $R+1$ sub-stencils, $S_m$, $m=1, \dots, R+1$.
Since the calculations described in Section \ref{sec:gp-weno} to obtain smoothness indicators need to
take place on every cell, there are $N_x$ operations involved in total, each of which has 
$M$ many operation counts at each $S_R$ level. Below, we estimate the total number of required operation counts
for Option 1 and Option 2. 

We first consider {\bf Option 1}. 
With a computer code with successive multiplications and additions for a dot product,
Eq. (\ref{eq:rec_fm}) involves $R+1$ multiplications and $R+1$ additions for
each $m=1, \dots, R+1$. This gives $2(R+1)$ operations for Eq. (\ref{eq:rec_fm}) for each $m$ and $l$.
Likewise, we also have $2(R+1)$ operations for Eq. (\ref{eq:alphas_1}) for each $m$ and $i$.
As a result, we see there are $2m(R+1)(l+i)$ operations involved on each $S_R$.
Since there are $N_x$ many $S_R$ calculations overall, the total number of
operations becomes $2m(R+1)(l+i)N_x$ per each time step, where
$m, i, l = 1, \dots R+1$. Assuming there are $M$ time steps required for the run,
we require $2m(R+1)(l+i)N_xM$ operations in total.

Let us now consider {\bf Option 2}. 
A major difference in this case is to realize the fact that Eq. (\ref{eq:Pvec})
can be pre-computed as soon as the grid is configured. 
This is because, unlike $f^{m}(x_l)$ in Eq. (\ref{eq:rec_fm}),
the vector $\bfP^{m}_i$ has no dependency on any local data but
only on the grid itself, whereby it can be pre-computed and saved for each $m$ and $i$
before evolving each simulation, and reused during the simulation evolutions.
The typical operation count for a $(R+1)$-dimensional vector and 
a $(R+1)\times(R+1)$-dimensional matrix multiplication with
successive multiplications and additions is found to be
$2(R+1)^2$. Hence we see there are $2im(R+1)^2$ total operations
that can be pre-computed and saved, $m,i=1, \dots, R+1$.
Here is an extra operation reduction available by realizing that Eq. (\ref{eq:Pvec})
is same for all $m$, thereby we have $2i(R+1)^2$ total operations for $i=1, \dots, R+1$.
Eq. (\ref{eq:alphas_2}) however, needs to be computed locally during
the run because it depends on the local data $\mathbf{G}_m$ that evolves in time.
For each $m$ and $i$, the dot product in Eq. (\ref{eq:alphas_2}) involves
$2(R+1)$ operations, totaling $2im(R+1)N_x$ per each time step, $m, i = 1, \dots R+1$.
As a result, we see there are $2i(R+1)^2$ operations initially, plus $2im(R+1)N_x$ during the run
per each time step. The simulation with $M$ time steps then involves 
$2i(R+1)^2 + 2im(R+1)N_xM = 2im(R+1)[R+1+N_xM]$ in total.

In comparison, we see there is a factor of two performance gain in {\bf Option 2} because
the ratio of the two options becomes
\bea
\frac{\mbox{Operation No. Option 1}} {\mbox{Operation No. Option 2}}
&&= \frac{2m(R+1)(l+i)N_xM}{2i(R+1)[R+1+mN_xM]}\\
&&= \frac{2 N_x M}{1+N_x M} \approx 2,
\eea
where we used $i,l,m=R+1$.

\bibliography{mybibfile}

\end{document}